\documentclass[12pt]{JHEP3}

\usepackage[title]{appendix}
\usepackage{amsmath,epsfig}
\usepackage{amssymb,amsfonts}
\usepackage{latexsym}
\usepackage{bbold}
\usepackage[latin1]{inputenc}
\usepackage{subeqnarray}
\usepackage{xcolor}
\let\normalcolor\relax
\usepackage{graphicx}
\usepackage{longtable}
\relax

\def\beq{\begin{equation}}
\def\eq{\end{equation}}
\def\bea{\begin{eqnarray}}
\def\ea{\end{eqnarray}}

\def\p{\partial}
\def\nn{\nonumber}
\def\half{\frac12}

\def\fdot{\dot{f}}
\def\Adot{\dot{A}}
\def\Addot{\ddot{A}}

\newcommand\fverb{\setbox\pippobox=\hbox\bgroup\verb}
\newcommand\fverbdo{\egroup\medskip\noindent%
                        \fbox{\unhbox\pippobox}\ }

\title{\huge Constraining Non-Relativistic RG Flows with Holography}
\author{\large Sera Cremonini$^{a}$,  Li Li$^{b,c,d}$, Kyle Ritchie$^{e}$, Yuezhang Tang$^{a}$\\
~\\
$^a$ Department of Physics, Lehigh University, Bethlehem, PA 18018, USA.\\
$^b$ CAS Key Laboratory of Theoretical Physics, Institute of Theoretical Physics, Chinese Academy of Sciences, P.O. Box 2735, Beijing 100190, China.\\
$^c$ School of Physical Sciences, University of Chinese Academy of Sciences, No.19A Yuquan Road, Beijing 100049, China.\\
$^d$ School of Fundamental Physics and Mathematical Sciences, Hangzhou Institute for Advanced Study, UCAS, Hangzhou 310024, China.\\
$^e$ Department of Physics and Astronomy, University of New Mexico, Albuquerque, NM 87131, USA.\\
E-mail: cremonini@lehigh.edu, liliphy@itp.ac.cn, kyleritchie@unm.edu, yut318@lehigh.edu
}

\abstract
{We examine non-relativistic holographic RG flows by working with 
Einstein-Maxwell-scalar theories which support geometries that break Lorentz invariance at some energy scale.
We adopt the superpotential formalism, which  
helps us characterize the radial flow in this setup and bring to light a number of generic features.
In particular, we identify several quantities that behave monotonically under RG flow. As an example, we show that the index of refraction is generically monotonic.
We also construct a combination of the superpotentials that flows monotonically 
in Einstein-scalar theories supporting non-relativistic solutions, 
and which reduces to the known c-function in the relativistic limit. 
Interestingly, such quantity also exhibits monotonicity in a variety of black hole solutions to the full 
Einstein-Maxwell-scalar theory, hinting at a deeper structure. 
Finally, we comment on the breakdown of such monotonicity conditions and on the relation to a candidate c-function obtained previously from entanglement entropy.}

\def\pdfannotlink{\pdfstartlink}

\begin{document}
\tableofcontents
\newpage

\section{Introduction}
\label{intro}

The basic question of how to organize the degrees of freedom of a quantum system as a function of energy scale has proven very difficult to address in a general context.
While  in relativistic quantum field theories (QFTs) powerful c-theorems (see \emph{e.g.} \cite{Zamol,Komargodski:2011vj,Freedman:1999gp}) provide a clear measure for the number of effective degrees
of freedom at a given energy -- through the existence of a \emph{monotonic} c-function -- they fail when Lorentz invariance is broken, with no clear pathway to extending them.
Yet it is an important fundamental question whether we can quantify the loss of  
information along renormalization group (RG) flow in a \emph{generic} quantum system, and in particular whether the loss of degrees of freedom is monotonic as we access lower scales.
Within the context of holography, this is related to the issue of how gravity encodes the process of integrating out field theoretic degrees of freedom, and how to geometrize RG flows in the presence of broken symmetries\footnote{For a discussion of the Wilsonian renormalization perspective in holographic theories, and in particular of the running of the couplings, emergent dynamics and the relation to the Hamilton-Jacobi functional equation, see \emph{e.g.}~\cite{deBoer:1999tgo,Faulkner:2010jy,Heemskerk:2010hk}.}.

While these questions remain challenging to answer, over the past decade we have seen many successful steps in holography to 
enlarge the universality class of quantum systems that can be probed using gravity duals.  
These include the construction of gravitational solutions  that realize geometrically a variety of emergent IR phases and RG flows connecting fixed points with different symmetries.
These are highly non-trivial examples of bulk evolution which offer a useful arena for probing basic properties of gradient flow and possible generalizations of c-theorems.
As a concrete example, solutions that interpolate between AdS vacua while traversing
intermediate non-relativistic scaling regimes (\emph{e.g.} described by hyperscaling violating, Lifshitz-like solutions~\cite{Bhattacharya:2014dea,Donos:2017ljs,Donos:2017sba,Hoyos:2020zeg})
provide geometric examples of emergent conformal symmetry and lead to basic questions about violations and generalizations of c-theorems\footnote{Other interesting cases are the holographic constructions with spontaneously broken translational symmetry that is relevant in the IR (without having any sources in the UV), which realize the idea of spontaneous crystallization~\cite{Rozali:2012es,Withers:2014sja,Cremonini:2017usb,Cai:2017qdz,Andrade:2017ghg,Cremonini:2018xgj,Ling:2019gjy}.}.

In this paper we examine holographic RG flows that don't assume Lorentz invariance at all scales, with the goal of identifying functions that may behave monotonically and 
thus capture the physics of a c-function.
For radial flows involving Lorentz invariant geometries, the monotonic c-function can be related to a ``superpotential" $W$ which allows one to recast 
the second order bulk gravity equations in the language of first order equations~\cite{Freedman:1999gp,Girardello:1998pd}. The latter are effectively RG flow equations. 
Indeed, the superpotential plays an important role in the Hamilton-Jacobi approach, which has a long history of describing holographic RG flow~\cite{deBoer:1999tgo,Papadimitriou:2007sj,Papadimitriou:2011qb}.
Recently, the authors of~\cite{Kiritsis:2016kog} examined a variety of relativistic flows arising in Einstein-scalar theories, relying on  the power of the superpotential 
formalism to obtain a general classification of gravitational solutions and a way to characterize the properties of the associated RG flows (see~\cite{Gursoy:2018umf,Bea:2018whf} for related studies at finite temperature and ~\cite{Kiritsis:2012ma} for an early study of non-relativitsic RG flows based on the quantum effective potential at finite temperature and density).

Motivated in part by~\cite{Kiritsis:2016kog}, here we work with non-relativistic solutions to Einstein-Maxwell-scalar theories and adopt the superpotential 
method developed in~\cite{Lindgren:2015lia}  to explore their structure.
As we will see, recasting the flow equations in terms of superpotentials will guide our intuition and 
help us identify certain quantities that are monotonic and may provide candidate c-functions.
We will also connect our analysis to that of~\cite{Cremonini:2013ipa}, who relied on 
the connection between entanglement entropy and c-theorems~\cite{Casini:2004bw,Myers:2010tj,Casini:2012ei} 
to examine a generalization of 
the standard c-function for spacetime geometries which spontaneously violate Lorentz invariance (see also \cite{Banerjee:2015coc,Chu:2019uoh} for more other attempts at generalized c-functions). 
Motivated by counterexamples to monotonicity of entanglement for broken Lorentz invariance~\cite{Swingle:2013zla}, 
\cite{Cremonini:2013ipa} explored how to constrain the geometry so that it could in principle yield monotonic flows. 
As we will see, the main points made in~\cite{Cremonini:2013ipa} can be recast naturally in the language of fake superpotentials. 
Finally, while some of our results are completely generic in Einstein-Maxwell-scalar theories  -- such as the monotonicity of the so-called index of refraction --  others will only hold in Einstein-scalar theories. Still, it is remarkable that even in the absence of relativistic symmetry some of the intuition from relativistic flows remains -- this is exactly what one would hope for from general intuition.

The outline of the paper is as follows. Section~\ref{Sec:setup} introduces our holographic setup and the fake superpotentials. Section~\ref{Sec:main} contains the main results of our analysis and examines the monotonic behavior of a variety of quantities relevant to describing the holographic RG flow. In Section~\ref{Sec:Examples} we illustrate with explicit examples some of our results on the radial flow of the superpotentials. Section~\ref{Sec:breakdown} examines the UV structure of the geometry to establish under which conditions monotonicity will break down. We conclude and summarize our discussions in Section~\ref{Sec:conclusion}.
In Appendix~\ref{App:extreme} we illustrate with a concrete example the statement that extremal geometries are always relativistic in the Einstein-scalar theories we examine. The  monotonicity of our superpotential $W$ in the Einstein-scalar theories is discussed in Appendix~\ref{App:W} .

\section{Holographic Setup and Background}\label{Sec:setup}

We examine Einstein-Maxwell-scalar theories of the form 
\begin{equation}
\label{appaction}
S =\frac{1}{2\kappa^2} \int d^{d+1}x \sqrt{-g} \left(R  - \dfrac{1}{2} (\partial \phi)^2 - V(\phi)-\frac{Z(\phi)}{4}F_{\mu\nu} F^{\mu\nu} \right) ,
\end{equation}
which from now on we refer to as EMD theories, and work
with backgrounds described by the ansatz
\begin{equation}\label{appmetric}
ds^2 = dr^2 + e^{2A(r)}(-f(r) dt^2 + d\vec{x}^2) \, ,\quad \phi=\phi(r), \quad A_\mu d x^\mu=A_t(r) dt \, .
\end{equation}
Thus, in the metric, the deviation from relativistic symmetry is captured by the function $f(r)$ not 
being a constant.
We will assume that the asymptotic UV boundary is at  $r\rightarrow\infty$. 
In order to rely on the standard holographic dictionary we take the geometry at the UV fixed point to be $AdS_{d+1}$. The conformal factor in the UV is then given by $A(r)=r/L$, with $L$ the AdS curvature scale, and 
$f(r\rightarrow\infty)$ approaches a constant\footnote{One might worry that the $AdS_2\times R^{d-1}$ geometry
does not seem to fall into the metric ansatz we are using. However, we note that  $AdS_2\times R^{d-1}$
 in our coordinate system is given by
\begin{equation}
ds^2 = dr^2 + e^{2A_0}(-e^{2 r} dt^2 + d\vec{x}^2)\,,\nn
\end{equation}
with $A_0$ a constant. 
After introducing $u=e^{r}$, we recover the more standard form of $AdS_2\times R^{d-1}$,
\begin{equation}
ds^2 =-e^{2A_0} u^2 dt^2+ \frac{1}{u^2}du^2 + e^{2A_0}d\vec{x}^2\,.\nn
\end{equation}
} there.

The equations of motion for these theories are given by
\bea
\label{Einstein1}
&& R_{\mu\nu} + \frac{Z}{2} \, F_{\mu\rho} F^{\rho}_{\;\;\nu} -  \half \p_\mu \phi \, \p_\nu \phi  +
\frac{g_{\mu\nu}}{2} \left[\half(\p \phi)^2 + V -R   +  \frac{Z}{4}  F^2  \right] = 0 \, , \nn \\
\label{scalar0}
&& \frac{1}{\sqrt{-g}}\, \p_\mu \left(\sqrt{-g}\,  \p^\mu \phi  \right) = \frac{1}{4} \frac{\p Z}{\p \phi} \, F^2 
+ \frac{\p V}{\p \phi}  \, , \nn \\
\label{gauge0}
&& \frac{1}{\sqrt{-g}} \, \p_\mu \left(\sqrt{-g} \, Z \, F^{\mu\nu}\right) = 0 \, .
\ea
Inserting the ansatz~\eqref{appmetric} into the above equations of motion, we obtain the following set of equations:
\begin{eqnarray}
\ddot{\phi}+\left(d\ddot{A}+\frac{\dot{f}}{2f}\right)\dot{\phi}-V_{\phi}+Z_{\phi}\frac{e^{-2A}}{2f}\dot{A_t}^2& = & 0\,,\\
\partial_r\left(e^{(d-2)A}\frac{Z\dot{A_t}}{\sqrt{f}}\right)&=&0\,,\\
\dot{\phi}^2+2(d-1)\left(\ddot{A}-\frac{\dot{f}}{2f}\right)&=&0\,,\\
\ddot{f}+\left(d\dot{A}-\frac{\dot{f}}{2f}\right)\dot{f}-Z e^{-2A}\dot{A_t}^2&=&0\,,\\
\ddot{A}+\left(d\dot{A}+\frac{\dot{f}}{2f}\right)\dot{A}+\frac{V}{d-1}+\frac{Z}{2(d-1)}\frac{e^{-2A}}{f}\dot{A_t}^2&=&0\,,
\end{eqnarray}
where the dot denotes the radial derivative, $V_{\phi}=\frac{d V}{d\phi}$ and $Z_{\phi}=\frac{d Z}{d\phi}$. Note that only two of the last three equations are independent.

By combining Einstein's equations with the mater equations of motion,
one obtains\footnote{This conserved quantity can be defined for a more general theory with broken U(1)  symmetry (see~\cite{Kiritsis:2015hoa}). It is possible to include additional field content, for example, axionic scalars.} the following radially independent quantity,
\begin{equation}\label{appQ}
\mathcal{Q}=-\frac{e^{d A}\dot{f}}{\sqrt{f}}+\frac{e^{(d-2)A}}{\sqrt{f}}Z A_t \dot{A_t}\,.
\end{equation}
The radially conserved combination $\mathcal{Q}$ is particularly useful because it can be used to connect IR to boundary data. We are interested in black brane solutions to this theory, for which there is a horizon at, say, $r = r_h$, with $f(r_h)=0$ and $A_t(r_h)=0$. One can then show that
\begin{equation}\label{appQboundary}
\mathcal{Q}(r_h)=-2\kappa^2 T\, s,\quad \mathcal{Q}(r\rightarrow\infty)=-2\kappa^2(\mathcal{E}+P-\mu\rho)\,.
\end{equation}
Here $T, s, \mathcal{E}, P,\mu,\rho$ are, respectively, the temperature, entropy density, energy density, pressure, chemical potential and charge density. 
The chemical potential is given by $\mu=A_t(r\rightarrow\infty)$, and the corresponding charge density is defined as
\begin{equation}\label{apprho}
\rho=\frac{1}{ 2\kappa^2}Z\sqrt{-g} F^{tr}|_{r\rightarrow\infty}=\frac{1}{ 2\kappa^2}e^{(d-2)A}\frac{Z\dot{A_t}}{\sqrt{f}}\,.
\end{equation}
Here the second equality is due to the fact that the combination $Z\sqrt{-g} F^{tr}$ is radially independent, and thus can be computed anywhere in the bulk.
Recall that $\mathcal{E}$ and $P$ can be obtained using holographic renormalization amended  by appropriate boundary terms, including the Gibbons-Hawking term for a well-defined Dirichlet variational principle and  surface counterterms for removing divergences.

Thus, we see that the conserved quantity $\mathcal{Q}$ just gives the expected thermodynamic relation\,\footnote{For more details and concrete examples, see the discussion of the consequence of the thermodynamic relation from $\mathcal{Q}$ in Einstein-scalar theories in~\cite{Li:2020spf}.}
\begin{equation}
 T\, s=\mathcal{E}+P-\mu\rho\,.
\end{equation}
Finally, for the extremal geometry describing the ground state of the theory, $Ts=0$ and thus  $\mathcal{Q}=0$.

\subsection{Superpotential Formalism}

The so-called fake superpotential method, which allows to describe solutions to second order equations of motion of a theory in terms of first order equations, 
has played a central role in the Hamilton-Jacobi approach as well as in the description of holographic RG flows. 
For the latter, work in the literature has been in the context of geometries which respect relativistic symmetries.
In this paper we plan to extend this analysis to our more general class of geometries, and explore the implications for monotonicity properties. 

To this end, we make use of the radial Hamiltonian formalism developed in \cite{Lindgren:2015lia}, where the authors also identified a 
fake superpotential $W$. Indeed, our model (\ref{appaction}) falls within the class of theories studied in \cite{Lindgren:2015lia}, 
where the superpotential $W$ was introduced\footnote{Note that~\cite{Lindgren:2015lia} used a different normalization for the scalars. We have also rescaled the superpotential by a factor of 1/2, \emph{i.e.} to match with the notation of \cite{Kiritsis:2016kog} one must send $W \rightarrow 2W$.} through
the action $ \mathcal{S}$ evaluated at a radial cutoff in the following way, 
\beq
 \mathcal{S} = -\frac{1}{2\kappa^2} \int d^d x \left[ e^{dA} \sqrt{f} \, W(A,\phi)  -2\kappa^2\rho A_t \right] \,.
\eq
Here  $A$, $f$,  $\phi$ and $A_t$ are, respectively, the metric components, scalar field and time component of the gauge field appearing in the 
ansatz (\ref{appmetric}). Finally, $\rho$ denotes  the background electric charge density.\footnote{Although we will not consider it here, the 
analysis of   \cite{Lindgren:2015lia} is also valid for backgrounds which include a magnetic field. Much of our later analysis can be generalized to include such cases.}

As shown in \cite{Lindgren:2015lia}, the superpotential and its derivatives obey
\bea
\label{appAdot}
\dot{A} &=&  - \frac{1}{2(d-1)} \, W \, ,  \\
\label{appfdot}
\frac{\dot{f}}{f} &=&  - \frac{1}{d-1} \, W_A \, ,  \\
\label{appphidot}
\dot\phi &=& W_\phi \, ,
\ea
where $W_A \equiv \partial_A W$ and similarly $W_\phi \equiv \partial_\phi W$.
Moreover, rearranging the equations of motion one can show that the superpotential obeys the first order differential equation
\beq
\label{appWeom}
\half W_\phi^2 - \frac{1}{4(d-1)} (d + \partial_A) W^2 = V_{eff}(\Phi) \, ,
\eq
where 
\beq
V_{eff}(\phi, A)= V (\phi)  + 2 Z^{-1} e^{-2(d-1)A}(\kappa^2\rho)^2  \, .
\eq
In the relativistic case corresponding to $f=1$ (and thus $W_A =0$), and in the absence of a vector field this reduces, as it should, to what was found in \cite{Kiritsis:2016kog}, 
\beq 
\half W_\phi^2 - \frac{d}{4(d-1)}  W^2 = V \, .
\eq

For completeness, we include here the null energy conditions (NEC) for the metric~\eqref{appmetric}, which are given by
\begin{eqnarray}\label{appNEC}
\frac{\fdot}{f} \Adot -2\Addot  \geq  0   \qquad \text{and} \qquad
d \Adot \fdot  +\ddot{f} - \frac{\fdot^2}{2f}  \geq  0       
\, ,
\end{eqnarray}
and can be read off immediately\footnote{To avoid getting confused over notation, we use $\tilde{f}$ and $\tilde{g}$ for the two functions of  \cite{Cremonini:2013ipa}
whose radial derivatives encode the two NECs. In terms of the notation of this section, we then have $\tilde{f} = - \frac{\Adot}{\sqrt{f}} = \frac{W}{2(d-1) \sqrt{f}}$  
and $\tilde{g} = \half e^{dA} \frac{\fdot}{\sqrt{f}} = - \sqrt{f} e^{dA} \frac{W_A}{2(d-1) }$.}
from 
\cite{Cremonini:2013ipa}. 
Note that for the theory~\eqref{appaction}, the equations of motion imply
\begin{equation}
d \Adot \fdot  +\ddot{f} - \frac{\fdot^2}{2f}=e^{-2A} Z  \dot{A_t}^2 \geqslant 0\,,
\end{equation}
and thus the second NEC condition is satisfied automatically.
Finally, expressed in terms of the superpotentials, the two conditions (\ref{appNEC}) become
\bea
\label{NEC1super}
\dot{W} + \frac{1}{2(d-1)} W \, W_A  & \geq & 0   \,,\\
\label{NEC2super}
d  \, W \, W_A - 2 (d-1) \dot{W}_A + W_A^2 & \geq & 0    \, .
\ea
We will return to these in the next section, when we comment on entanglement entropy and connect to the results of \cite{Cremonini:2013ipa}.

\section{Monotonicity Conditions and Constraints}\label{Sec:main}

Having setup the holographic system we are going to work with, we ask whether there are generic features that appear in the
structure of the solutions, as the energy scale of the dual field theory is varied. In particular, we inspect the radial flow 
of various quantities including the superpotentials $W$ and $W_A$, with the goal of identifying possible monotonic features.
Indeed, we will see that there are certain special quantities which vary monotonically as they approach the IR from the UV fixed point, and which can be used to characterize the flow.

\subsection{Monotonicity of warp factor and  Raychauduri's equation}\label{sec:warp}

We start by examining the properties of the warp factor $A(r)$, whose radial derivative is related to the superpotential $W$ through (\ref{appAdot}).
As we will show, in our construction the sign of $W$ can be easily fixed. Indeed, this can be seen by  
adopting a new radial coordinate $z$,
\begin{equation}\label{appnewcord}
ds^2 =\frac{1}{z^2H(z)} dz^2+\frac{1}{z^2}\left(-F(z)dt^2 + d\vec{x}^2\right) \,,
\end{equation}
in terms of which the asymptotic UV boundary is now located at $z=0$. 
To recover the original form~\eqref{appmetric} of the metric we have been using, one needs the following identification,
\begin{equation}
dr =-\frac{1}{z\sqrt{H(z)}} dz,\quad e^{2A(r)}=\frac{1}{z^2},\quad f(r)=F(z) \,.
\end{equation}
It then immediately follows that
\begin{equation}
\label{Adot}
\dot{A}(r)=\frac{d A(r)}{dr} =\frac{dz}{dr}\frac{dA(r)}{dz}=\frac{dz}{dr}\frac{d(-\text{ln}(z))}{dz}=-z\sqrt{H(z)}\frac{d(-\text{ln}(z))}{dz}=\sqrt{H(z)}\geqslant 0\,,
\end{equation}
showing that $A(r)$ is monotonic and fixing the sign of the superpotential, 
\beq
\label{Wsign}
W \leq 0 \, .
\eq

This result can also be obtained using the Raychauduri's equation. Let's consider the congruence of in-going null geodesics that is given by
\begin{equation}
n^a=\frac{e^{-2A}}{f}\left(\frac{\partial}{\partial t}\right)^a-\frac{e^{-A}}{\sqrt{f}}\left(\frac{\partial}{\partial r}\right)^a\,.
\end{equation}
Then the expansion parameter $\theta$ of the congruence is obtained by
\begin{equation}\label{appexpansion}
\theta=g^{ab}\nabla_a n_b=-\frac{(d-1)e^{-A}}{\sqrt{f}}\dot{A}\,.
\end{equation}
Imposing the NEC, from Raychauduri's equation one obtains
\begin{equation}
\frac{d\theta}{d\lambda} \leqslant-\frac{1}{d-1} \, \theta^2\,,
\end{equation}
with $\lambda$ the affine-parameter.
Therefore, if the expansion parameter takes a negative value $\theta_0$ at any point on a geodesic in the congruence, the expansion will be negative until it arrives at, if it exists, a caustic at which $\theta$ becomes negative infinity and jumps from $-\infty$ to $+\infty$. 
For the background geometry~\eqref{appmetric}, near the UV boundary we have $A(r)=r/L$ and $f(r\rightarrow\infty)$ approaches a positive constant.
Thus, one obtains from~\eqref{appexpansion} that $\theta\sim-\frac{d-1}{L}e^{-r/L}<0$ near the UV. 
If there is no caustic for the geometry~\eqref{appmetric} from the IR to the UV, $\theta$ can not increase along the null geodesic, and thus we must have $\dot{A}\geqslant 0$.

We point out that the author of~\cite{Bousso:1999xy} proposed to use congruences of null geodesics as probes 
 for the sampling of the holographic encoding of information in gravitational theories.
This idea has been realized by~\cite{Sahakian:1999bd}, in which the expansion parameter of the null congruence
was used to extract the information encoded holographically in the geometry. The key point is the requirement for the convergence of in-going null geodesics~\cite{Sahakian:1999bd} 
\begin{equation}\label{apptheta}
\theta\leqslant 0\,,
\end{equation}
along the radial direction. This condition agrees with our earlier result \eqref{Adot}. What's more, from our arguments based on the coordinate system~\eqref{appnewcord}, one can show that indeed there is no caustic outside the horizon for our background~\eqref{appmetric}.
Thus, it is safe to conclude that $\dot{A}\geqslant 0$. We also stress that the monotonicity of the warp factor $A(r)$ is purely geometrical,  \emph{i.e.} it involves no use of any fields equation. We also showed that if the Einstein's equations holds with the matter stress energy tensor satisfying the NEC, then $A(r)$ should be monotonic in the absence of caustic for the geometry.

Finally, we note that the authors of~\cite{Kolekar:2018chf} examined holographic RG flows for geometries that have an infrared $AdS_2$ throat region. They introduced a holographic c-function to describe the number of active degrees of freedom along the RG flow to the IR $AdS_2$ fixed point\,\footnote{We would like to thank K. Narayan for bringing this work to our attention.}. Motivated by their discussion, we define the entropy density associated with the cutoff surface at a fixed radial position $r$,
\begin{equation}\label{effentropy}
s_{c}=\frac{2\pi}{\kappa^2}e^{(d-1) A(r)}\,,
\end{equation}
where we are working with the background geometry~\eqref{appmetric}.
Since we have shown that $\dot{A} \geq 0$, it is clear that  $s_c$ increases monotonically as a function of $r$, suggesting that the active degrees of freedom decrease along the RG flow from the UV to IR. We stress that our result is quite generic and is independent of the IR geometry.

\subsection{Refraction index monotonicity in EMD theories}\label{sec:index}
The next quantity which we are interested in examining is the \emph{refraction index} $n$, which was introduced in~\cite{Gubser:2009gp} to describe the relative speed of propagation of light-like signals (for an early discussion of mechanisms to obtain a varying speed of light on a probe brane see \cite{Kiritsis:1999tx}). To our knowledge, very little is known about possible universal properties of $n$ in general holographic setups, and the most  extensive discussions in holography to date can be found in~\cite{Donos:2017ljs,Donos:2017sba,Hoyos:2020zeg,Gauntlett:2018vhk,Arav:2018njv}. In particular, the authors of~\cite{Donos:2017ljs,Donos:2017sba} examined the UV/IR behavior of $n$ (which describes the renormalization of length scales in the system) and how it varies as a function of the deformation parameter which controls the holographic boomerang RG flows. 
In the supersymmetric model studied in~\cite{Gauntlett:2018vhk}, the monotonicity of the index of refraction (as a function of the radial coordinate) was shown to follow from the BPS equations, and it was suggested that this should be a generic feature. 
More recently, it was shown in~\cite{Hoyos:2020zeg} that in the corresponding holographic constructions $n$ flows monotonically from the UV to the IR.
Here we would like to better understand the behavior of $n$ and, in particular, to identify 
the potential origin of such monotonicity, if it indeed generic.
 
For our metric~\eqref{appmetric} the index of refraction at a particular radial location has the form $n(r) = \sqrt{f}$ and its radial derivative, from which one can determine its monotonicity, is given by
\begin{equation}
\dot{n} = \half \frac{\dot{f}}{\sqrt{f}}\,.
\end{equation}
Using~\eqref{appQ} and~\eqref{appQboundary} one finds that
\begin{equation}
\dot{n} = \half \frac{\dot{f}}{\sqrt{f}}=-\half\mathcal{Q} \,e^{-d A}+\kappa^2 e^{-d A}A_t\,\rho=\kappa^2e^{-d A}(T\,s+A_t\,\rho)\,,
\end{equation}
where we have used the definition of charge density~\eqref{apprho}.

Note that in the simple case of Einstein-scalar theory without a gauge field ($A_t =0$), we have
\begin{equation}
\dot{n} =\kappa^2 e^{-d A}T\,s\geqslant 0\,,
\end{equation}
implying that $n$ is necessarily monotonic. 
The lower limit $\dot{n}=0 $, corresponding to $n$ being constant along the flow, is particularly interesting.
It describes geometries which are extremal, since $T=0$ for such solutions.
Indeed, for pure Einstein-scalar theories the extremality condition combined with (\ref{appQ}) implies
\begin{equation}\label{appextremeQ}
\mathcal{Q}=-\frac{e^{d A}\dot{f}}{\sqrt{f}}=0 \Rightarrow f(r)=constant\,.
\end{equation}
This means that \emph{extremal geometries in Einstein-scalar theory are relativistic}
independently of which potential $V(\phi)$ one chooses\,\footnote{Some concrete examples are illustrated in Appendix~\ref{App:extreme}.}, \emph{i.e.} they are described by a metric of the form
\begin{equation}
ds^2 = dr^2 + e^{2A(r)}(-dt^2 + d\vec{x}^2) \,,
\end{equation}
where for convenience we have set $f(r)=1$. 
This is not  true for Einstein-Maxwell-scalar theories due to the contribution from the second term of~\eqref{appQ}.
Note that there is an alternative way to reach this result. Recall that in the analysis of~\cite{Cremonini:2013ipa}  the deviation from a relativistic metric was captured by 
a non-zero value of the function $g$ introduced there, which was useful because it allowed to express the NEC in a simple way, $g^\prime \geq 0$.
As it turns out, for Einstein-scalar theories 
 the quantity (\ref{appQ}) is is essentially the non-relativistic function $g$ of~\cite{Cremonini:2013ipa}, 
$$ \mathcal{Q}=-2 g \, .$$
Thus, we immediately see that any $\mathcal{Q}=0$
geometry supported by such theories will have to be relativistic. 
From this we conclude that in Einstein-scalar theories $n=constant$ corresponds to extremal geometries which are relativistic.

More generally, in the presence of a gauge field 
one immediately has $\dot{n}\geqslant 0$ as long as $A_t\,\rho\geqslant 0$. Thus, if the latter 
condition holds we have once again that $n$ is monotonic along the radial direction. 
To prove that indeed $A_t\,\rho\geqslant 0$ for charged black holes, we take 
without loss of generality  $\rho>0$. Then one observes from~\eqref{apprho} that
\begin{equation}
\dot{A_t}={ 2\kappa^2} e^{(2-d)A}\frac{\sqrt{f}}{Z}\rho\geqslant 0 \, ,
\end{equation}
implying that $A_t$ increases monotonically with increasing radius in general. Using the condition $A_t(r_h)=0$, one immediately finds that
\begin{equation}
A_t(r)>0,\quad (r>r_h)\,,
\end{equation}
with a positive chemical potential $\mu=A_t(r\rightarrow\infty)$.
Thus, we have $A_t\,\rho\geqslant 0$ for the charged case, telling us that we always have
\beq
\dot{n} \geq 0  \, .
\eq
Thus, \emph{the refraction index $n$ is always monotonic for Einstein-Maxwell-scalar theories} of the form (\ref{appaction}). 
It would be interesting to explore to what extent the monotonicity of $n$ is a completely general feature, or if it no longer holds in the presence of a more complicated matter sector.
The boomerang flows of~\cite{Donos:2017ljs,Donos:2017sba,Hoyos:2020zeg} would be a particularly intriguing examples to examine (note that  
the refraction index was indeed monotonic in the flows considered in~\cite{Hoyos:2020zeg}).
Finally, note that in terms of the superpotential we have
\beq
\dot{n} = \half \frac{\fdot}{\sqrt{f}} = - \frac{\sqrt{f}}{2(d-1)} W_A \, ,
\eq
and thus the monotonicity of $n$ is connected to $\dot{f}$ or $W_A$ having a definite sign. More specifically, it implies that  
\beq
\dot{f} \geq 0
 \qquad \text{or equivalently} \qquad W_A \leq 0 \,.
\eq

\subsection{Superpotentials and monotonicity conditions}
\label{SuperSubsection}

After analysing the behavior of the two metric components $A$ and $f$, which are both monotonic functions of the radial coordinate, we are able to fix the signs of the superpotentials in EMD theories, 
\beq
\label{signs}
W \leq 0 \, , \qquad W_A \leq 0 \, .
\eq
Next, we inspect in more detail how the superpotentials behave under RG flow, in the hope of identifying additional quantities (other than the refractive index $n$) which may be monotonic and potentially play the role of a c-function -- providing a quantitative measure for the varying number of degrees of freedom along the flow.
For relativistic backgrounds (for which $W_A=0$) the c-function is related to the superpotential $W$ in a straightforward way, and the monotonicity of the former follows immediately from that of the latter.
In particular, a monotonic superpotential is guaranteed by the fact that its radial derivative obeys $\dot W \sim \dot \phi^2 \geq 0$, and this is essentially at the core of the existence of a c-theorem.
Motivated by these observations, we want to examine how to generalize this story to non-relativistic geometries.

Recall that in the relativistic case, the beta function for the scalar field in Einstein-scalar theories was related to the superpotential in a simple way \cite{Kiritsis:2016kog}, through 
\beq
\label{betarel}
\beta(\phi) = - 2(d-1) \frac{W_\phi}{W} \, .
\eq
In the case of non-relativistic backgrounds arising in the EMD theories (\ref{appaction}), this construction can be easily extended.
In particular, we denote the QFT energy scale by
\beq
E = E_0 \sqrt{f} e^A \, ,
\eq 
and define the beta function via
\beq
\beta(\phi) \equiv \frac{d\phi}{d\log E} = E \frac{d\phi}{d E} \,.
\eq
We then find that (\ref{betarel}) generalizes to 
\beq
\frac{d\phi}{d\log E} = \frac{\dot\phi}{\dot{A} + \frac{\dot{f}}{2f} }  \quad \Rightarrow \quad \beta(\phi) = -2 (d-1) \frac{W_\phi}{W + W_A} \, .
\eq
We now see clearly that the flow is controlled by the behavior of both $A$ and $f$, as expected.
Moreover, in the relativistic case $W_A=0$ and one recovers the result of \cite{Kiritsis:2016kog}, a simple check on the analysis.
Finally, using (\ref{signs}) we conclude that for our EMD theory the sign of $\beta$ is entirely determined by the sign of $W_\phi$, an intriguing feature which is unexpected.

Appropriately rearranging the equations of motion for our EMD model, we can extract the radial derivatives 
of the superpotentials.
In particular, from the equation of motion for $A$ we find that 
\beq
\label{EOMA}
\frac{dW}{dr} = \dot\phi^2 - \frac{W W_A}{2(d-1)} \, ,
\eq
which tells us that $W$ is \emph{not} generically monotonic for 
backgrounds that are non-relativistic, since $W W_A \geq 0$ and therefore the two terms on the right hand side compete with each other.\,\footnote{The breakdown of monotonicity of W at finite temperature in the Einstein-scalar theories is discussed in Appendix~\ref{App:W}.}
In the relativistic case we recover $\dot W = \dot\phi^2 \geq 0$ as anticipated. 
For EMD theories, the second superpotential $W_A$ obeys 
\beq
\frac{dW_A}{dr} = \frac{d}{2(d-1)} W_A \left(W + \frac{W_A}{d} \right)  - (d-1) Z(\phi) e^{-2A}  \frac{\dot A_t^2}{f} \, ,
\eq
where again in general we see a competition between the first term containing $W$ and $W_A$, which is always positive due to (\ref{signs}), and the gauge field term, which is always negative (note that $Z(\phi)>0$). 
However, in the special case of Einstein-scalar theories (for which $A_t=0$) we have the stronger result 
\beq
\label{WAsimple}
\frac{dW_A}{dr} = \frac{d}{2(d-1)} W_A \left(W + \frac{W_A}{d} \right)  \geq 0 \, ,
\eq
implying that $W_A$ is always monotonic.

Thus, so far we have identified three quantities  that, under appropriate conditions, behave monotonically under RG flow: the warp factor $A$, the index of refraction $n$ in EMD theories, and the superpotential $W_A$ in Einstein-scalar theories. 
Note, however, that both $n$ and $W_A$ are trivial in the relativistic limit --  the former becomes a constant and the latter vanishes -- and therefore are not general enough to capture 
the expected relativistic behavior and the physics of a sensible c-function.
We would like to do better, and look for a different quantity that may be monotonic and in addition reduce to the known relativistic result.
To this end, we consider first the simpler case of Einstein-scalar theory, and take the following 
combination of (\ref{EOMA}) and (\ref{WAsimple}),
\beq
\label{EOMcombined}
\frac{d}{dr} \left(W + \frac{1}{d} W_A\right) = \frac{1}{2d(d-1)}W_A^2 +\dot\phi^2 \geq 0 \, .
\eq
Thus, we immediately see that the combination 
\beq
W + \frac{1}{d} W_A 
\eq
is always monotonic in Einstein-scalar theories.
Furthermore, it reduces to the correct c-function in the presence of relativistic symmetry, and is therefore a better candidate than either $n$ or $W_A$.

Going back now to the general EMD theory, 
(\ref{EOMcombined}) is modified as follows, 
\beq
\label{EOMcombinedGaugeField}
\frac{d}{dr} \left(W + \frac{1}{d} W_A\right) = \frac{1}{2\, d(d-1)}W_A^2 +\dot\phi^2 - \frac{(d-1)  }{d} Z(\phi) \, e^{-2A} \,  \frac{\dot A_t^2}{f} \, ,
\eq
where again we see a clear competition between the first two terms and the last one -- 
the negative contribution from the gauge field term spoils the monotonicity condition.
We still have a monotonic flow when the right hand side of (\ref{EOMcombinedGaugeField}) has a definite sign from the UV to the IR,
but the physical interpretation of this condition is not clear. 
Still, we have computed the quantity (\ref{EOMcombinedGaugeField}) for many known solutions in the literature, and found several instances in which the flow is monotonic.
It is surprising and intriguing that this seems to be the norm rather than the exception.
We have devoted Section~\ref{Sec:Examples} to a detailed discussion of these cases,  and in Section~\ref{Sec:breakdown} we discuss the conditions under which the monotonicity breaks down.

\subsection{Connections to entanglement entropy and c-functions}
Next, we would like to consider the constraints from NEC (\ref{appNEC}) and entanglement entropy, revisiting 
the arguments of \cite{Cremonini:2013ipa}.
Recall that a candidate c-function was suggested by the authors of~\cite{Myers:2012ed}:
\beq
\label{candidate}
c_d \equiv \beta_d \frac{\ell^{d-1}}{H^{d-2}} \, \frac{\partial S_{EE}}{\partial \ell} \, ,
\eq
where $S_{EE}$ denotes the \emph{holographic} entanglement entropy for a strip-shaped region of width $\ell$, $\beta_d$ is a numerical factor and the size of the plane $H \gg \ell$ can be thought of as an infrared regulator.
The holographic entanglement entropy for such a region can then be shown to be given by~\cite{Cremonini:2013ipa}
\begin{equation}
S_{EE} = \frac{4\pi H^{d-2}}{\kappa^2} \int_{r_m}^{r_c} dr \frac{e^{(d-2)A}}{\sqrt{1- e^{-2(d-1)(A-A(r_m))}}} \,,
\end{equation}
with the width of the strip
\begin{equation}
\ell = 2 \int_{r_m}^\infty dr \frac{e^{-A}}{\sqrt{e^{2(d-1)(A-A(r_m))}-1}}\,,
\end{equation}
where $r_c$ is a fixed UV cutoff and $r_m$ is the turning point of the bulk minimal surface (\emph{i.e.} the minimal radius which it reaches in the bulk).

The criterion for a monotonic $c$-function is that $d c_d/d\ell$ should have a definite sign, and in particular we expect it to be be negative (or zero for trivial flows) in this setup. 
The authors of \cite{Myers:2012ed}, who worked with relativistic geometries, showed that 
$d \ell/d r_m\le 0$ for minimal surfaces under the assumption that $\dot{A}=\frac{d A}{d r}\ge 0$.
Note in particular that indeed $\dot{A}\geqslant 0$ as shown in section~\ref{sec:warp}.
So one will recover the standard Wilsonian RG intuition provided that $d c_d/d r_m \geq 0$.

Following the strategy of~\cite{Myers:2012ed}, the running of $c_d$ as a function of $r_m$ was computed by~\cite{Cremonini:2013ipa}, and found to be given by
\beq
\label{appdotc}
\frac{d c_d}{d r_m} = -\gamma\, \dot{A}(r_m)
\int_0^{\ell} dx \frac{\ddot{A}}{\dot{A}^2} =\frac{\gamma}{2}\,\dot{A}(r_m)\int_0^{\ell} dx \frac{1}{\dot{A}^2}\left[\left(\frac{\fdot \Adot }{f} -2\Addot\right)-\frac{\fdot \Adot }{f} \right]\,,
\eq
with $\gamma$ a positive numerical factor. 
In the second equality, the integrand inside the round parentheses is non-negative due to NEC, see~\eqref{appNEC}. 
However, NEC is not enough to ensure that (\ref{appdotc}) has a definite sign.
The condition $\fdot \Adot\leqslant 0$ would in principle guarantee that we always have ${d c_d}/{d r_m}\geqslant0$ and hence a monotonic
flow for the c-function (it would be a sufficient condition, but not necessary). 
Indeed, this was noted in \cite{Cremonini:2013ipa}, who discussed restrictions on (the UV values of) the metric functions $A$ and $f$ that ensured ${d c_d}/{d r_m}\geqslant0$ along the radial flow.
However, in general one can not have $\fdot \Adot\leqslant 0$. 
In fact, we have just seen that for the Einstein-Maxwell-scalar theories~\eqref{appaction} we are considering, $\dot{f}\dot{A}\geqslant 0$ instead (this follows from the fact that $\dot{A} \geq 0$ and 
$\dot{n} \geq 0$).

Another way to state this is to note that the monotonicity of $c_d$ is affected
 by whether $\Addot \propto \dot{W}$ changes sign, and from the discussion around (\ref{EOMA}) we know that $W$ is not generically monotonic in EMD theories. Thus, we shouldn't expect $c_d$ to be monotonic, unless special restrictions on the geometry are imposed.
Here we are restating some of the results of \cite{Cremonini:2013ipa}, but now interpreted from the point of view of the superpotential.
We also note that while in the relativistic case $\Addot \leq 0$
 (which ensures that $\frac{d c_d}{d r_m} \geq 0$), for non-relativistic backgrounds one can in principle have $\Addot \geq 0$. According to 
(\ref{appdotc}), this would correspond to a monotonic flow ``in the wrong direction," with the candidate c-function decreasing towards the UV.

To summarize, one has a monotonic $c_d$ provided $\int_0^{\ell} dx \frac{\ddot{A}}{\dot{A}^2}$ in~\eqref{appdotc} has a definite sign. 
What this discussion also indicates is that this particular candidate c-function, derived from the entanglement entropy of an infinite strip, may not be the best quantity to capture the 
number of degrees of freedom at different energy scales. 
A better quantity to use, at least in the case of Einstein-scalar theories, may be $W+\frac{1}{d}W_A$, which is indeed monotonic along the radial flow.
Perhaps the index of refraction $n$, which is monotonic generically in these theories, is also a viable measure to adopt.

\subsection{Monotonicity of the energy scale}

Before proceeding to examining specific examples, we comment briefly on the energy scale that we will adopt to describe our holographic RG flows. 
To discuss RG flow one needs to introduce an energy scale, which in holography should relate to the radial coordinate in the bulk. 
A fundamental formulation of the relation between scales on the boundary and bulk physics is still not fully understood, and in particular how to 
define a proper energy scale that can capture the behavior of the dual boundary field theory is still an open question.

In the relativistic context, corresponding to setting $f=1$ in the metric (\ref{appmetric}), the energy scale has been taken to be $E = E_0 \,  e^A$ with $E_0$ a constant. 
To take into account the non-relativistic features of the flows we are examining, we propose the following scale, 
\beq 
E = E_0 \sqrt{f} e^A  \, .
\eq
It should vary monotonically with the radial coordinate, and reduce to the one used in the relativistic case (which clearly it does). 
To determine whether it flows monotonically, we must inspect the sign of its radial derivative,
\beq\label{eqmu}
\dot E  = E  \left(\Adot + \frac{\fdot}{2f} \right) = - \frac{E}{2(d-1)} (W+W_A) \, .
\eq
Recall that we have shown that both  $\dot{A}$ and $\dot{f}$ are non-negative, or alternatively in terms of the superpotentials
\beq
\label{WWAsigns}
W \leq 0 \, , \qquad  W_A \leq 0 \, .
\eq
Using (\ref{WWAsigns}) we see that $\dot E $ has a definite sign, and thus $E$ is indeed a monotonic function of $r$. Moreover, it increases
as one approaches the UV boundary.
This is expected and consistent with the holographic picture, and in particular the UV-IR relation.
Finally, note that in our present study the precise relation between the energy scale $E$ and the radial position is not important, because we are only interested in the monotonicity condition of the RG flow. 
Moreover, it is sufficient to consider the flow as a function of the radial coordinate, since the energy scale $E$ has been shown to be increase monotonically along the radial coordinate from IR to UV.

\section{Examples}
\label{Sec:Examples}

Next, we will examine the behavior of the superpotentials $W$ and $W_A$ as they flow from the UV to the IR using explicit analytical black hole solutions.
In particular, we will test the monotonicity properties of the quantity 
$W+\frac{1}{d}W_A$ using certain black hole solutions to  
EMD theories. 
We will find, surprisingly, that the monotonicity of $W+\frac{1}{d}W_A$ is stronger than naively apparent from (\ref{EOMcombinedGaugeField}).

\subsection{Supergravity solutions in $AdS_4$}
We consider three special cases of maximal gauge supergravity in four dimensions, which were first obtained from a more general black hole with four U(1) charges in~\cite{Cvetic:1999xp}. 
The precise form of the solutions is summarized in Table~\ref{tab:sg}, with the background ansatz taking the following form,
\begin{equation}
\label{sgmetric}
ds^2_4=\frac{d\bar{r}^2}{e^{2\bar{A}(\bar{r})} \bar{f}{(\bar{r})} }  + e^{2\bar{A}(\bar{r})} (-\bar{f}(\bar{r})d t^2+d\vec{x}^2 ) \, , \quad \phi = \bar{\phi}(\bar{r}) \,,\quad  A_\mu dx^\mu=\bar{A}_t(\bar{r})dt\,.
\end{equation}
The corresponding temperature is given by
\begin{equation}
T=\frac{\bar{f}'(\bar{r}_h)e^{2\bar{A}(\bar{r}_h)}}{4\pi}\,,
\end{equation}
with $\bar{r}_h$ the location of the event horizon.  
See~\cite{Kiritsis:2015oxa} for a more detailed discussion of the thermodynamic and spectral properties.

\begin{table}
\centering
\caption{\label{tab:sg} Analytic solutions from supergravity in AdS$_4$.}
\renewcommand{\arraystretch}{2.5}
\setlength\doublerulesep{0.2pt}
\begin{scriptsize}
\begin{tabular}{|c|c|c|c|}
  \hline\hline
  & one-charge & two-charge & three-charge \\

  \hline
  $V(\phi)$ & $-6\cosh(\phi/\sqrt{3})$ & $-2(\cosh\phi+2)$ & $-6\cosh(\phi/\sqrt{3})$\\

  \hline
  $Z(\phi)$ & $e^{\sqrt{3}\phi}$ & $e^\phi$ & $e^{\phi/\sqrt{3}}$\\

  \hline
  $e^{2A}$ & $\bar{r}^{3/2}(\bar{r}+Q)^{1/2}$ & $\bar{r}(\bar{r}+Q)$ & $\bar{r}^{1/2}(\bar{r}+Q)^{3/2}$\\

  \hline
  $f$ & $1-\dfrac{\bar{r}_h^2(\bar{r}_h+Q)}{\bar{r}^2(\bar{r}+Q)}$ & $1-\dfrac{\bar{r}_h(\bar{r}_h+Q)^2}{\bar{r}(\bar{r}+Q)^2}$
  & $1-\dfrac{(\bar{r}_h+Q)^3}{(\bar{r}+Q)^3}$\\

  \hline
  $A_t$ & $\dfrac{\sqrt{Q}\bar{r}_h}{\sqrt{\bar{r}_h+Q}}\Bigl(1-\dfrac{\bar{r}_h+Q}{\bar{r}+Q}\Bigr)$ & $\sqrt{2Q\bar{r}_h}\Bigl(1-\dfrac{\bar{r}_h+Q}{\bar{r}+Q}\Bigr)$ & $\sqrt{3Q(\bar{r}_h+Q)}\Bigl(1-\dfrac{\bar{r}_h+Q}{\bar{r}+Q}\Bigr)$\\

  \hline
  $\phi$ & $\dfrac{\sqrt{3}}{2}\ln\Bigl(1+\dfrac{Q}{\bar{r}}\Bigr)$ & $\ln\Bigl(1+\dfrac{Q}{\bar{r}}\Bigr)$ & $\dfrac{\sqrt{3}}{2}\ln\Bigl(1+\dfrac{Q}{\bar{r}}\Bigr)$\\

  \hline
  $T$ & $\dfrac{(3\bar{r}_h+2Q)\sqrt{\bar{r}_h}}{4\pi\sqrt{\bar{r}_h+Q}}$ & $\dfrac{3\bar{r}_h+Q}{4\pi}$
  & $\dfrac{3\sqrt{\bar{r}_h(\bar{r}_h+Q)}}{4\pi}$\\

  \hline
  $\mu$ & $\dfrac{\sqrt{Q}\bar{r}_h}{\sqrt{\bar{r}_h+Q}}$ & $\sqrt{2Q\bar{r}_h}$ & $\sqrt{3Q(\bar{r}_h+Q)}$\\

  \hline\hline
\end{tabular}
\end{scriptsize}
\end{table}

Before proceeding, we connect the coordinate system of~\eqref{sgmetric} with the one we used in our discussion,~\eqref{appmetric}. It is easy to see that the two coordinate systems are related by
\begin{equation}
dr=\frac{d\bar{r}}{e^{\bar{A}(\bar{r})} \sqrt{\bar{f}{(\bar{r})}}}\,, 
\end{equation}
with $\{A(r), f(r), \phi(r), A_t(r)\}=\{\bar{A}(\bar{r}), \bar{f}(\bar{r}),\bar{\phi}(\bar{r}), \bar{A}_t(\bar{r})\}$. 
Armed with this, we can obtain the corresponding superpotential quantities in the $\bar{r}$-coordinate,
\begin{eqnarray}
W & = & -4 \dot{A}(r) =-4 e^{\bar{A}(\bar{r})}\sqrt{\bar{f}(\bar{r})}\,\bar{A}'(\bar{r})\,,\\
W_A & =&  -2\dot{f}/f=-2  e^{\bar{A}(\bar{r})}\sqrt{\bar{f}(\bar{r})}\,\frac{\bar{f}'(\bar{r})}{\bar{f}(\bar{r})}\,,\\
W+\frac{1}{3}W_A & =&-4 \dot{A}(r)-\frac{2}{3}\dot{f}/f= -2  e^{\bar{A}(\bar{r})}\sqrt{\bar{f}(\bar{r})} \left (2\bar{A}'(\bar{r})+\frac{1}{3}\frac{\bar{f}'(\bar{r})}{\bar{f}(\bar{r})}\right)\,.
\end{eqnarray}
We can now check the monotonicity behavior of the superpotential quantities we discussed in the previous section using these explicit supergravity solutions. We shall work in the grand canonical ensemble with the chemical potential $\mu$ fixed to be unity. 
It is straightforward to check that both the refraction index $n$ and the warp factor $A$ are increasing monotonically towards the UV. 

\begin{figure}[ht!]
\begin{center}
\includegraphics[width=2.8in]{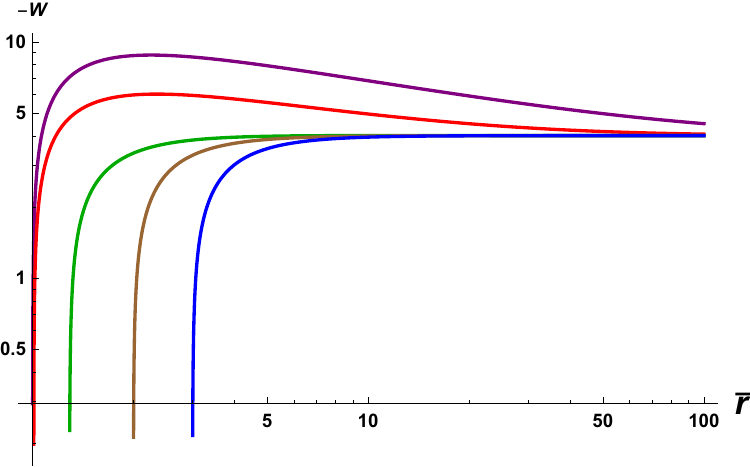}\quad
\includegraphics[width=2.8in]{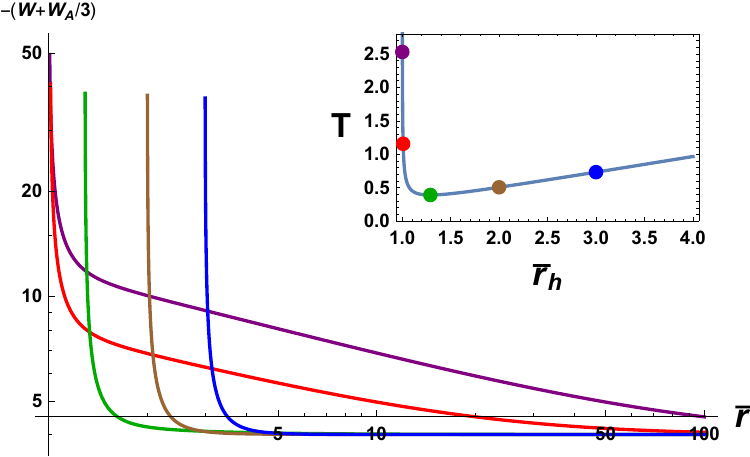}
\caption{The superpotentials with respect to the radial coordinate $\bar{r}$ for the one-charge black hole at different temperatures.  The inset shows the temperature as a function of $\bar{r}_h$ with $\bar{r}_h>1$. There are two branches of solutions  for a given temperature above the minimum temperature $T_m=\sqrt{3/2}/\pi$ at $\bar{r}_h=\sqrt{5/3}$ (green). The superpotential $W$ (left panel) is monotonic in the branch with large $\bar{r}_h$, while non-monotonic in the small $\bar{r}_h$ branch. In contrast, the quantity $W+\frac{1}{3} W_A$ (right  panel) is always monotonic as a function of $\bar{r}$. 
We have worked in units with $\mu=1$.}
\label{fig:onecharge}
\end{center}
\end{figure}

\begin{figure}[ht!]
\begin{center}
\includegraphics[width=2.8in]{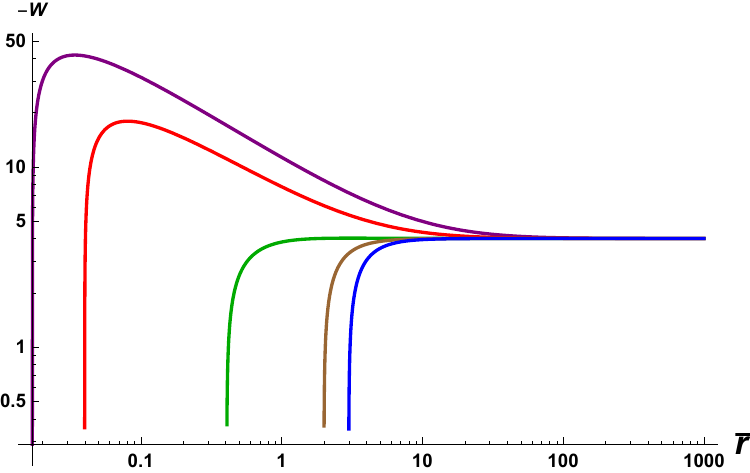}\quad
\includegraphics[width=2.8in]{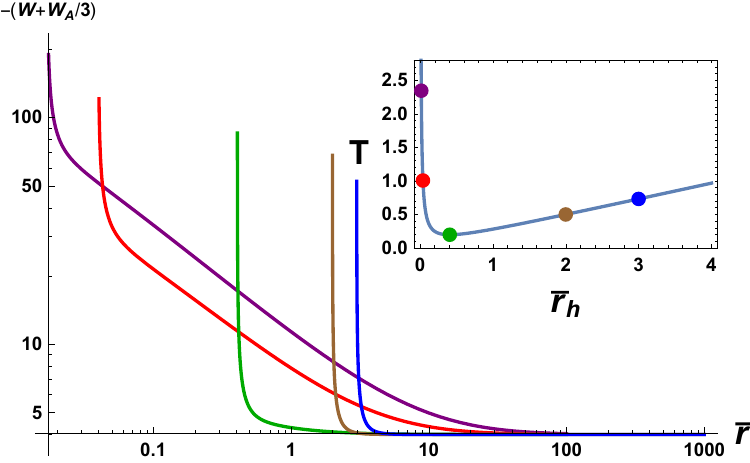}
\caption{The superpotentials with respect to the radial coordinate $\bar{r}$ for the two-charge black hole at different temperatures.  The inset shows the temperature as a function of $\bar{r}_h$ with $\bar{r}_h>0$. There are two branches of solutions  for a given temperature above the minimum temperature $T_m=\sqrt{3/2}/(2\pi)$ at $\bar{r}_h=\sqrt{1/6}$ (green). The superpotential $W$ (left panel)  is monotonic in the branch with large $\bar{r}_h$, while non-monotonic in the small $\bar{r}_h$ branch. In contrast, the quantity $W+\frac{1}{3} W_A$ (right panel) is always monotonic as a function of $\bar{r}$. We have worked in units with $\mu=1$.}
\label{fig:twocharge}
\end{center}
\end{figure}

\begin{figure}[ht!]
\begin{center}
\includegraphics[width=2.8in]{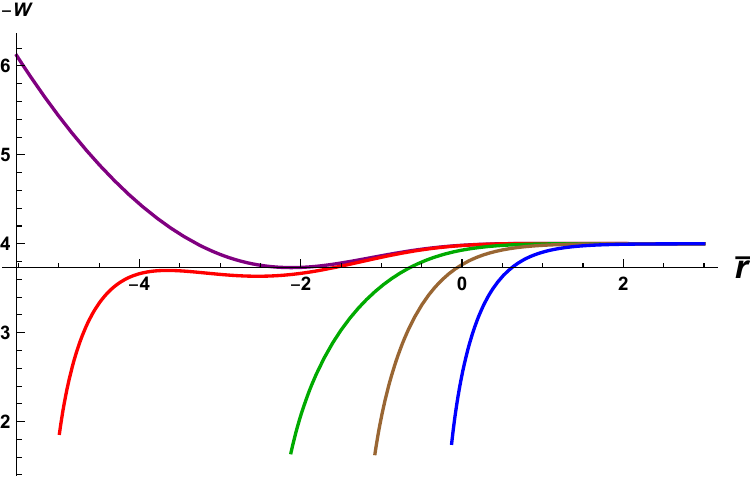}\quad
\includegraphics[width=2.8in]{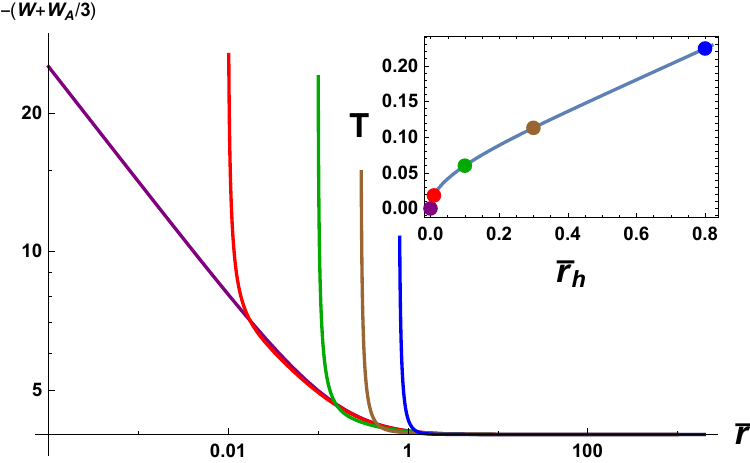}
\caption{The superpotentials with respect to the radial coordinate $\bar{r}$ for the three-charge black hole at different temperatures.  The inset shows the temperature as a function of $\bar{r}_h$ with $\bar{r}_h\geqslant0$ (the extremal case corresponds to the purple curve with $\bar{r}_h=0$). The extremal IR geometry is semi-local and is conformal to $AdS_2\times R^2$. The superpotential $W$ (left panel) is not monotonic at low temperatures, but becomes monotonic when the temperature is increased. The combination $W+\frac{1}{3} W_A$ (right panel) is always monotonic as a function of $\bar{r}$. We have worked in units with $\mu=1$.}
\label{fig:threecharge}
\end{center}
\end{figure}

The superpotential $W$ and the combination $W+\frac{1}{3}W_A$ for the one-charge, two-charge and three-charge black hole solutions are shown in Figures~\ref{fig:onecharge}, \ref{fig:twocharge} and~\ref{fig:threecharge}, respectively. 
In the first two cases we see that there are two branches of black hole solutions for a given temperature: a branch with a small horizon radius $\bar{r}_h$, and the other one corresponding to a large value of $\bar{r}_h$. 
Moreover, there is a minimum temperature $T_m$ below which no black hole solutions exist. 
On the other hand, in the case of the three-charge black hole there is only one branch of solutions, and the extremal IR geometry is conformal to $AdS_2\times R^2$. 
As one can see, the one-charge and two-charge black holes share similar features.
In particular,  $W$ (shown in the left plots) is monotonic in the branch with large $\bar{r}_h$, while it is non-monotonic in the small $\bar{r}_h$ branch. For the three-charge case, $W$  behaves non-monotonically at low temperatures, but becomes monotonic when the temperature is increased sufficiently. 
The crucial feature, however, is that in all these cases, $W+\frac{1}{3}W_A$ is  monotonic as a function of $\bar{r}$. 
More precisely, it increases monotonically towards the UV.

\subsection{Black Brane Solutions in the STU Model}\label{sec:STU}

Next, we examine a class of analytical five-dimensional black brane solutions that arise from a particular 
limit of the STU model in maximally supersymmetric gauged supergravity\footnote{The Kaluza-Klein reduction of Type IIB supergravity on $S^5$ can be consistently truncated to the lowest-mass modes lying in a single five-dimensional multiplet, resulting in five-dimensional maximally supersymmetric gauged supergravity. The STU model is the consistent truncation of the full theory that includes the metric, three gauge fields and two scalars.}. 
These solutions, which were studied in~\cite{DeWolfe:2012uv}, correspond to a limit of the STU model in which 
two of the three charges are taken to be equal to each other. In this truncation of the STU model, only one scalar field is turned on, along with two gauge fields and gravity. 
Moreover,  the one-charge and two-charge black brane solutions in this theory are supported by just one gauge field, so that the effective lagrangian is of the same form we are considering in~\eqref{appaction}, with the scalar potential 
\beq
V(\phi) =- \frac{8}{L^2} e^{\phi/\sqrt{6}} - \frac{4}{L^2} e^{-2\phi/\sqrt{6}} \, , 
\eq 
and the couplings of the scalar to the gauge field\,\footnote{We have rescaled the gauge 
fields of~\cite{DeWolfe:2012uv} via $a_\mu\rightarrow a_\mu/2, A_\mu\rightarrow A_\mu/2\sqrt{2}$ in order to use the standard normalization for $Z$, \emph{i.e.} $Z(\phi=0)=1$.}
 given by 
\beq
Z(\phi) = e^{-4\phi/\sqrt{6}} \quad \text{one-charge case} \, , \qquad  
Z(\phi) = e^{2\phi/\sqrt{6}} \quad \text{two-charge case} \, . 
\eq  
The background ansatz is of the form
\begin{equation}
ds^2_5=\frac{e^{2\bar{B}(\bar{r})} d\bar{r}^2}{ \bar{f}{(\bar{r})} }  + e^{2\bar{A}(\bar{r})} (-\bar{f}(\bar{r})d t^2+d\vec{x}^2 ) \, , \quad \phi = \bar{\phi}(\bar{r}) \,,\quad  A_\mu dx^\mu=\bar{A}_t(\bar{r})dt\,,
\end{equation}
and the temperature is given by
\begin{equation}
T=\frac{\bar{f}'(\bar{r}_h)}{4\pi}e^{\bar{A}(\bar{r}_h)-\bar{B}(\bar{r}_h)}\,,
\end{equation}
with $\bar{r}_h$ the location of the event horizon. 
To connect to the coordinate system of~\eqref{appmetric} we use
\begin{equation}
dr=\frac{e^{\bar{B}(\bar{r})}  d\bar{r}}{\sqrt{\bar{f}{(\bar{r})}}}\,.
\end{equation}
For the one-charge black hole the background is given by 
\bea\label{oneSTU}
 &&  \bar A (\bar r) = \ln\frac{\bar r}{L} + \frac{1}{6}\ln \Bigl( 1+ \frac{Q^2}{\bar r^2}\Bigr) ,  \quad
\bar B (\bar r) = -\ln \frac{\bar r}{L} - \frac{1}{3}\ln \left( 1+ \frac{Q^2}{\bar r^2}\right) ,  \\
&& \bar f(\bar r) = 1 - \frac{r_h^2(\bar r_h^2 + Q^2)}{r^2(\bar r^2 + Q^2)}, \;
\bar \phi(\bar r) =-\sqrt{\frac{2}{3}} \ln \left(1+ \frac{Q^2}{\bar r^2} \right), \;
\bar A_t(\bar r) =\mu \left(1- \frac{\bar r_h^2 + Q^2}{\bar r^2 + Q^2} \right)\,, \nonumber
\ea
where the chemical potential\footnote{Because of the gauge field rescaling, our chemical potential is twice that of 
\cite{DeWolfe:2012uv}.} $\mu$ and the temperature are 
\beq
\mu = \frac{Q \bar{r}_h}{L \sqrt{\bar{r}_h^2 + Q^2}}, \qquad T =\frac{2\bar{r}_h^2 + Q^2}{2 \pi L^2 \sqrt{\bar{r}_h^2 + Q^2}}  \,.
\eq
\begin{figure}[ht!]
\begin{center}
\includegraphics[width=2.8in]{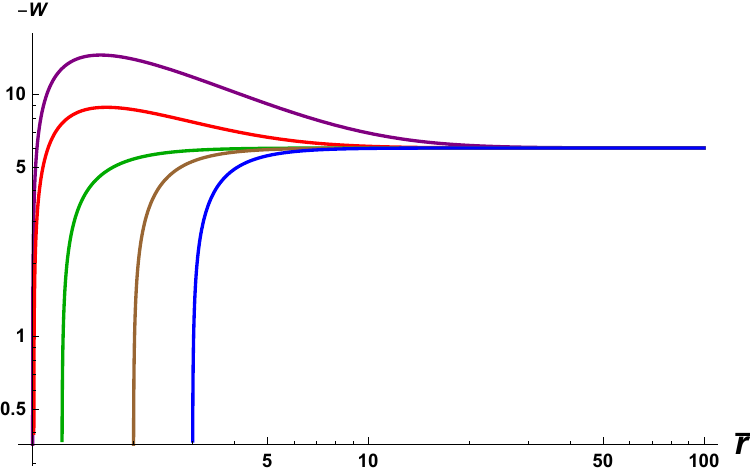}\quad
\includegraphics[width=2.8in]{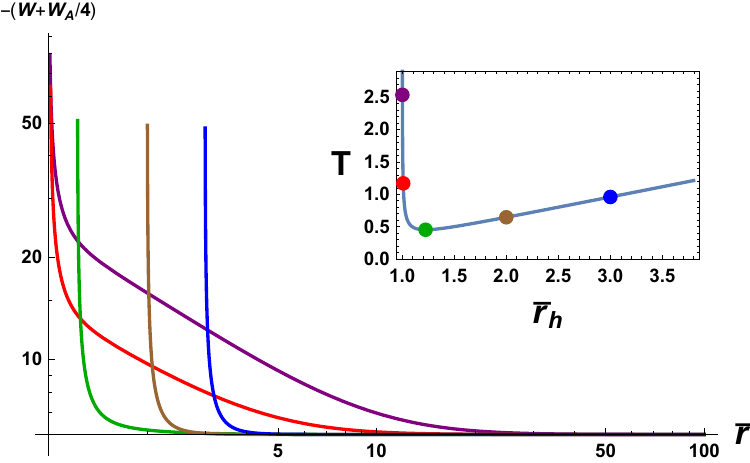}
\caption{The superpotentials with respect to the radial coordinate $\bar{r}$ for the one-charge black hole in the STU model.  The inset shows the temperature as a function of $\bar{r}_h$ with $\bar{r}_h>1$. There are two branches of solutions  for a given temperature above the minimum temperature $T_m=\sqrt{2}/\pi$ at $\bar{r}_h=\sqrt{3/2}$ (green). The superpotential $W$ (left panel)  is monotonic in the branch with large $\bar{r}_h$, while non-monotonic in the small $\bar{r}_h$ branch. In contrast, the quantity $W+\frac{1}{4}W_A$ (right panel) is always monotonic as a function of $\bar{r}$. We have worked in units with $\mu=1$ and $L=1$.}
\label{fig:onechargeSTU}
\end{center}
\end{figure}
\begin{figure}[ht!]
\begin{center}
\includegraphics[width=2.8in]{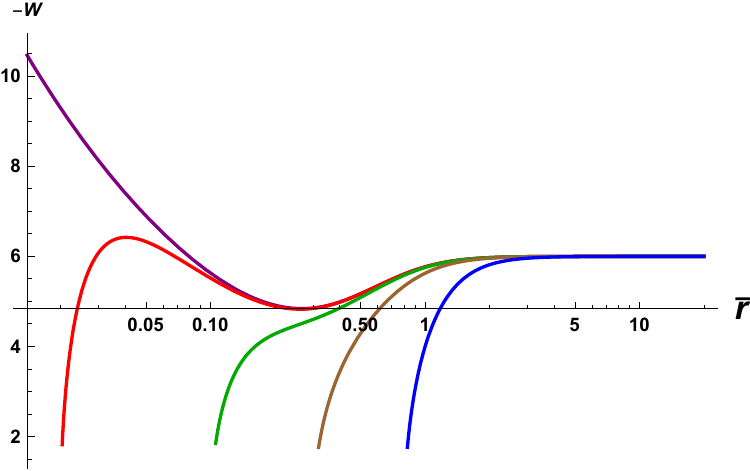}\quad
\includegraphics[width=2.8in]{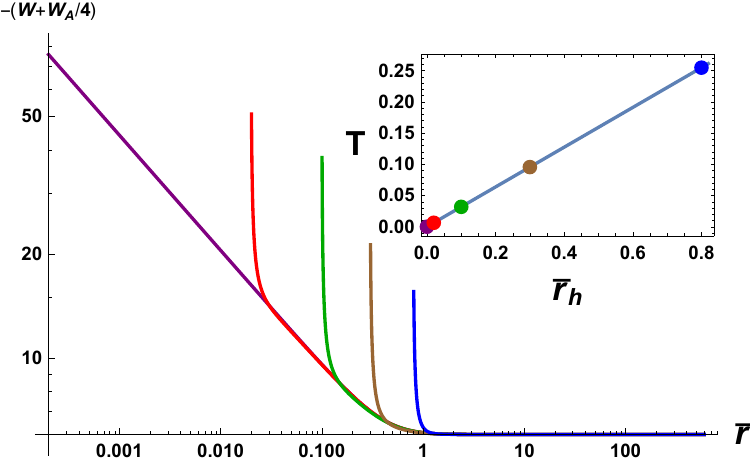}
\caption{The superpotentials with respect to the radial coordinate $\bar{r}$ for the two-charge black hole in the STU model at different temperatures.  The inset shows the temperature as a function of $\bar{r}_h$ with $\bar{r}_h\geqslant0$ (the extremal case corresponds to the purple curve with $\bar{r}_h=0$). The superpotential $W$ (left panel)  is not monotonic at low temperatures, but it becomes monotonic when the temperature is increased. $W+\frac{1}{4}W_A$ (right panel) is always monotonic as a function of $\bar{r}$. We have worked in units with $\mu=1$ and $L=1$.}
\label{fig:twochargeSTU}
\end{center}
\end{figure}
For the 2-charge black hole instead we have
\bea\label{twoSTU}
 &&  \bar A (\bar r) = \ln\frac{\bar r}{L} + \frac{1}{3}\ln \Bigl( 1+ \frac{Q^2}{\bar r^2}\Bigr) ,  \quad
\bar B (\bar r) = -\ln \frac{\bar r}{L} - \frac{2}{3}\ln \left( 1+ \frac{Q^2}{\bar r^2}\right) ,  \\
&& \bar f(\bar r) = 1 - \frac{(\bar r_H^2 + Q^2)^2}{(\bar r^2 + Q^2)^2}, \;
\bar \phi(\bar r) =\sqrt{\frac{2}{3}} \ln \left(1+ \frac{Q^2}{\bar r^2} \right), \;
\bar A_t(\bar r) =\frac{\sqrt{2}Q}{L} \left(1- \frac{\bar r_H^2 + Q^2}{\bar r^2 + Q^2} \right) ,\nonumber
\ea
with the chemical potential\footnote{Once again, our chemical potential needs to be divided by $2\sqrt{2}$ to agree with \cite{DeWolfe:2012uv}.} $\mu$ and temperature given by
\beq
\mu = \frac{\sqrt{2}Q }{L}, \qquad T =\frac{r_H}{\pi L^2 }  \,.
\eq

For both of these solutions we have tested the monotonicity properties of the combination 
\beq
W + \frac{1}{4} W_A 
=- \left(6 \dot{A} + \frac{3}{4} \frac{\dot f}{f}\right)
= - 3 e^{-\bar B}\sqrt{\bar f} \left(2 {\bar A^\prime} + \frac{1}{4} \frac{\bar f^\prime}{\bar f}\right) ,
\eq
where primes denote derivatives with respect to the radial coordinate $\bar{r}$.
The corresponding plots for the one-charge and two-charge cases are given in Figure~\ref{fig:onechargeSTU} and Figure~\ref{fig:twochargeSTU}, respectively.
We see that both cases lead to a monotonic flow for the combination $W + \frac{1}{4}W_A$, although the superpotential $W$ does not behave monotonically. This provides yet another example of a non-trivial class of solutions supported by gauge fields for which the combination $W + \frac{1}{d}W_A$ behaves monotonically.
Next, we examine the generic UV expansion for the fields in our theory, and identify a set of conditions that can guide us to determine whether we should expect a breakdown of the monotonicity.

\section{The Breakdown of Monotonicity and UV Expansions} \label{Sec:breakdown}

By analyzing the UV structure of the geometry, we can be more quantitative about 
the 
breakdown of the monotonicity of $W + \frac{1}{d} W_A$, which we recall 
obeys the following relation,
\beq\label{eqWWA}
\frac{d}{dr} \left(W + \frac{1}{d} W_A\right) = \frac{1}{2\, d(d-1)}W_A^2 +\dot\phi^2 - \frac{(d-1) Z \, e^{-2A}  }{d}  \frac{\dot A_t^2}{f}\, .
\eq
In particular, in order to determine if $W + \frac{1}{d} W_A$ increases monotonically towards the UV, a good starting point is to check its behavior near the AdS boundary. 
Without loss of generality, near the AdS boundary we parameterize the scalar coupling and scalar potential as follows,
\begin{equation}\label{expansionZV}
Z(\phi)=1+\alpha \phi+\dots,\quad V(\phi)=-\frac{d(d-1)}{L^2}+\frac{1}{2}m^2\phi^2+\dots\,,
\end{equation}
where we have chosen $\phi\rightarrow 0$ at the UV fixed point for convenience, and the parameter $m^2$ is the mass of the scalar field.

The asymptotic expansions near the AdS boundary $r\rightarrow \infty$ are schematically given by
\begin{eqnarray}
\phi(r)& = & \phi_s\, e^{-\Delta_{-} r/L}+\dots+\phi_v\, e^{-\Delta_{+} r/L}+\dots\,,\\
A_t(r)&=&\mu-\frac{2\kappa^2 L\rho}{d-2}e^{-(d-2)r/L}+\dots\,,\\
f(r)& = & 1- M e^{-d\,r/L}+\dots,\\
A(r)&=&\frac{r}{L}+\dots\,,
\end{eqnarray}
with\footnote{In the special case with $m^2L^2=-d^2/4$, corresponding to the UV BF bound with $\Delta_{-}=\Delta_{+}=d/2$, one has $\phi(r) = \frac{r}{L} \phi_s\, e^{-\frac{d}{2}\frac{r}{L}}+\phi_v\, e^{-\frac{d}{2} \frac{r}{L}}+\dots$. This case dose not change our general discussion below. A concrete example is the STU mode discussed in section~\ref{sec:STU}, where both black brane solutions~\eqref{oneSTU} and~\eqref{twoSTU} have no sources, \emph{i.e.} $\phi_s=0$.} 
 $\Delta_{\pm}=(d\pm\sqrt{d^2+4m^2L^2})/2$. 
 The constant $M$ can be fixed by the horizon data, $T s$, as well as the charge density via the radially conserved quantity $\mathcal{Q}$~\eqref{appQ}.

We are interested in identifying cases for which the quantity $W +\frac{1}{d}W_A$ may not increase monotonically.
Let's first consider the case without the scalar source term, for which 
we obtain from~\eqref{eqWWA} that
\begin{equation}\label{nosource}
\frac{d}{dr} \left(W + \frac{1}{d} W_A\right) =w_M M^2 e^{-2d\,r/L} +w_v \phi_v^2 e^{-2\Delta_{+}r/L}-w_\rho\rho^2 e^{-2(d-1)r/L}+\dots \, ,
\end{equation}
with $w_M, w_v, w_\rho$ positive constants. Note that the first term is smaller than the third term. 
What this implies\footnote{There is also a critical case with $\Delta_{+}=d-1$, for which one needs to compare the coefficients of the last two terms of~\eqref{nosource}.}
 is that $\frac{d}{dr} \left(W +  \frac{1}{d} W_A \right)$ can be negative when $\Delta_{+}>d-1$, which is equivalent to having 
$m^2L^2>1-d$.

For the case with a non-vanishing source $\phi_s$ we have
\begin{eqnarray}
\frac{d}{dr} \left(W + \frac{1}{d} W_A\right) =w_M M^2 e^{-2d\,r/L} +w_s \phi_s^2 e^{-2\Delta_{-}r/L}-w_\rho \rho^2 e^{-2(d-1)r/L}+\dots\,,
 \end{eqnarray}
where $w_s$ is a positive constants. Note that $\Delta_{-}=(d-\sqrt{d^2+4m^2L^2})/2\leqslant d/2$, immediately telling us that $\Delta_{-}<d-1$ for generic $d>2$. 
Thus, the second positive term itself is sufficient to guarantee that  $\frac{d}{dr} \left(W +  \frac{1}{d} W_A \right)$ is positive near the AdS boundary.

\begin{figure}[ht!]
\begin{center}
\includegraphics[width=3.8in]{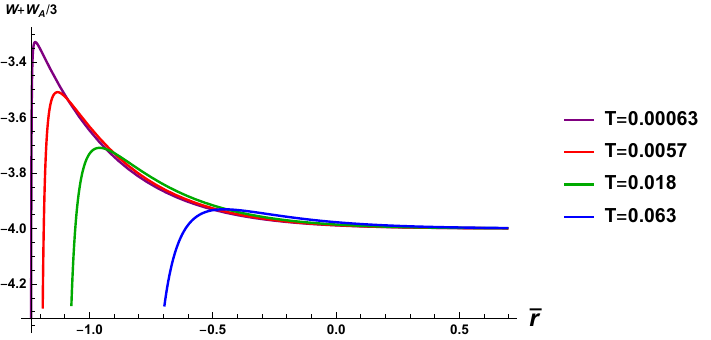}
\caption{The superpotential $W+\frac{1}{3}W_A$ as a function of the radial coordinate $\bar{r}$ for the four dimensional Reissner-Nordstr\"{o}m black hole. It behaves non-monotonically with respect to $\bar{r}$. We have worked in the units with $\mu=1$.}
\label{fig:RN}
\end{center}
\end{figure}

To summarize what we found, one can violate the monotonically increasing behavior of $W + \frac{1}{d} W_A$ in the following case\footnote{For the black brane at finite temperature one finds $\frac{d}{dr} \left(W +  \frac{1}{d} W_A \right)=\frac{2(d-1)}{d}\frac{1}{(r-r_h)^2}+\dots$ near the horizon. Thus, it is manifest that in the special case~\eqref{special} the quantity $W +  \frac{1}{d} W_A$ is non-monotonic along the radial flow.}
\begin{eqnarray}\label{special}
\phi_s=0,\quad m^2L^2>1-d \,.
\end{eqnarray}
As an extreme example, we check the Reissner-Nordstr\"{o}m black hole where $\phi=0$. According to the discussion above, $W+\frac{1}{d} W_A$ will decrease towards the UV near the AdS boundary.
In terms of the coordinate system~\eqref{sgmetric}, one has
\begin{eqnarray}
\bar{f}(\bar{r})=1-\frac{\bar{r}_h^3}{\bar{r}^3}+\frac{\mu^2\bar{r}_h^2}{4\bar{r}^4}\left(1-\frac{\bar{r}}{\bar{r}_h}\right),\quad e^{2\bar{A}(\bar{r})}=\bar{r}^2\,,
\quad \bar{A}_t(\bar{r})=\mu\left(1-\frac{\bar{r}_h}{\bar{r}}\right)\,,
\end{eqnarray}
where we have chosen $V=-6$ and $Z=1$ in a four dimensional spacetime ($d=3$). 
We display the corresponding behavior of  $W+\frac{1}{3} W_A$ as a function of the radial coordinate $\bar{r}$ in Figure~\ref{fig:RN}.
For a given temperature, it first increases and then decreases as $\bar{r}$ is increased.

Apart from the special case~\eqref{special}, one can in principle have  $W +  \frac{1}{d}W_A$ 
increase monotonically towards the UV. 
For all the cases with a non-vanishing scalar source we have checked, this is 
indeed the case.

\section{Summary of Results and Conclusions}\label{Sec:conclusion}

The issue of how to quantify the number of effective degrees of freedom of a system as a function of 
its energy scale is a longstanding one, and it's particularly challenging when one can not rely on 
a high degree of symmetry. As an example, the c-theorem hasn't been extended yet to theories that are not Lorentz invariant at all energy scales.
It is still reasonable to expect, however, that we should be able to track the loss of information or the loss of degrees of freedom in an arbitrary quantum system, and that said loss should be monotonic as the energy scale varies, at least under some appropriate set of restrictions.

In this paper we have attempted to make further steps in this direction within the framework 
of holographic RG flows, by identifying a variety of monotonic quantities in the class of EMD theories (\ref{appaction}), 
working with non-relativistic geometries of the form (\ref{appmetric}). 
In our analysis it has been particularly useful to adopt the superpotential formalism, which 
was worked out by~\cite{Lindgren:2015lia} for a broad class of theories including ours.
More specifically, for non-relativistic metrics of the form 
\beq
ds^2 = dr^2 + e^{2A(r)}(-f(r) dt^2 + d\vec{x}^2) \, ,
\eq
one can introduce the two 
superpotentials 
$ W = -2(d-1) \dot A$ and $W_A=-(d-1) \dot{f}/f$, which can then be utilized, along with 
$W_\phi = \dot\phi$, to recast the second order bulk equations into a first order differential equation, analogous to a flow equation. 
It is the function $f$, or equivalently the second superpotential $W_A$, which encodes the deviation from Lorentz invariance in this setup ($W_A=0$ for relativistic symmetry).

For relativistic theories it is well known that the superpotential $W$, which varies monotonically 
with the holographic coordinate, plays the role of the $c$-function. Thus, its behavior is intrinsically useful in order to understand basic properties of holographic RG flows.
Here we have examined to what extent one can identify quantities that are monotonic in our more general setup, where the background geometry is non-relativistic and $W$ is no longer the 
only superpotential which is directly associated with the geometry.
We summarize here the main results of our work:
\begin{enumerate}
  \item Extremal geometries arising in Einstein-scalar theories of the form $\mathcal{L} = R -\half
  (\partial \phi)^2 -V(\phi)$ are relativistic. To have a non-relativistic extremal geometry, one has to include, for example, a gauge field.
  \item  In the EMD theory (\ref{appaction}) the warp factor $A$ and the metric function $f$ 
  are both monotonic, \emph{i.e.} $\dot{A}\geqslant 0$ and $\dot{f}\geqslant 0$. 
The first one guarantees a monotonic effective entropy density $s_c$~\eqref{effentropy}. The latter implies that the index of refraction $n$ is also monotonic.
  \item Both $W_A$ and the combination $W + \frac{1}{d} W_A $ are monotonic in Einstein-scalar theories. The superpotential $W$, on the other hand, is only monotonic when $\dot\phi^2 - \frac{W W_A}{2(d-1)} >0$. Thus, in this class of theories the quantity $W + \frac{1}{d} W_A $ offers a possible generalization of the relativistic c-function (given by $W$) and reduces to it when $W_A=0$, as expected.
\item Although the special combination $W + \frac{1}{d} W_A $ is \emph{not} generically monotonic in the full EMD theory (\ref{appaction}),  it does vary in a monotonic fashion for many known black hole solutions. That its monotonicity is somewhat robust is illustrated with a number of examples in Section~\ref{Sec:Examples}. The breakdown of this behavior can be quantified by examining the UV structure of the geometry, as shown in Section~\ref{Sec:breakdown}.
  \item
  Using the conditions $\dot{f}\geqslant 0$ and $\dot{A}\geqslant 0$ we obtain $W_A\leqslant 0$ and $W\leqslant0$. Then~\eqref{eqmu} tells us that in the EMD theory (\ref{appaction}) the energy scale $E$ varies monotonically with the radial coordinate.
  In particular, $E$ increases as one moves to the UV boundary, which is consistent with the holographic picture and in particular the UV-IR relation, although a fundamental formulation of the relation between scale on the boundary and bulk physics is yet unknown.
\end{enumerate}

We have also connected to the analysis of~\cite{Cremonini:2013ipa} and recast some of the 
 properties of the c-function $c_d$ obtained from holographic entanglement entropy in terms of the superpotential $W$. 
Unlike in the relativistic case, the NEC (\ref{appNEC})
is not enough to ensure that $c_d$ is monotonic,
as discussed in \cite{Cremonini:2013ipa}. 
The monotonicity of the c-function is guaranteed provided~\eqref{appdotc} has a definite sign -- 
more specifically, when $\ddot{A}\sim - \dot W$ has a definite sign, \emph{i.e.} $W$ is monotonic.
When $\ddot{A}\leqslant 0$,  $W$ is everywhere an increasing function of the holographic coordinate and the c-function increases towards the UV, 
$d c_d/d r_m \geq 0$. On the other hand, when $\ddot{A}\geqslant 0$ 
(taking into account 
the geodesic condition $\dot{A}\geqslant 0$) one finds that $d c_d/d r_m \leqslant 0$.  
This describes a monotonic RG flow but with the opposite sign, with the c-function flowing in the wrong direction. However, we stress 
that the condition $\ddot{A}\geqslant 0$ can only be realized with non-relativistic geometries
($\ddot{A}\leqslant 0$ in the relativistic case).
Finally, even when $W$ is not monotonic -- which will generically be the case -- one still has a monotonic $c_d$ provided the integrand in~\eqref{appdotc} has a definite sign.

The fact that the c-function $c_d$ obtained from entanglement entropy is only 
sensitive to $\ddot{A} \sim - \dot W$ already tells us that it may not be the most appropriate quantity to consider, when thinking of generalizations of the relativistic c-function.
In particular, we expect both $W$ and $W_A$ to play a crucial role in characterizing the flow.
This intuition is confirmed by the structure of the $\beta-$function which we have derived in 
Section 3, $\beta(\phi) = -2(d-1) \frac{W_\phi}{W+W_A}$, which generalizes the relativistic 
version of \cite{Kiritsis:2016kog}.
It is interesting that the special combination $W + \frac{1}{d} W_A $, which is \emph{always} monotonic in Einstein-scalar theories but not generically in EMD theories, exhibits monotonicity 
for a variety of charged black hole solutions to (\ref{appaction}).
Thus, this behavior is somewhat robust, even when $W$ itself is 
clearly not monotonic. 
We would like to better understand if this is a sign of a deeper structure, emerging at least in certain ranges of parameter space or temperature, or whether it is simply accidental. 
It would also be interesting to extend our analysis to theories that can support the  
boomerang RG flows of~\cite{Donos:2017ljs,Donos:2017sba,Hoyos:2020zeg}, which require a more complicated matter content than 
the EMD model examined here.

All in all, for the Einstein-scalar theory we have stronger constraints on the monotonicity of RG flows. 
We have identified additional quantities that behave monotonically in the more general framework of EMD theories, in particular the warp factor (or, equivalently, the effective entropy density $s_c$) and the refraction index $n$.
However, the latter vanishes for relativistic geometries and thus is not a particularly viable 
candidate for a generalized c-function, which should reduce to the known Lorentz invariant result in the appropriate limit.
It is then natural to ask what is the appropriate way to quantify the changing number of degrees of freedom -- which quantity, if any, can be used to generalize the notion of a c-function?
While this issue remains open, 
our analysis clearly shows that even in non-relativistic theories 
one can extract holographic features that are generic, and identify 
geometric quantities that behave monotonically under RG flow.
There is still a question about their physical interpretation and fundamental origin, and of whether they have a natural interpretation in the language of quantities such as entanglement entropy.

We should also note that in the non-relativistic case one can define a slightly different superpotential $\tilde{W}$, 
related to ours through $\tilde{W}=W/\sqrt{f}$. 
Using the equations of motion, it is straightforward to check that $\frac{d\tilde{W}}{dr}=\frac{\dot{\phi}^2}{\sqrt{f}}\geqslant 0$ for Einstein-scalar theories, showing that $\tilde{W}$ is monotonic, as discussed by \cite{Gursoy:2018umf} using a slightly
 different coordinate system from ours. 
 It would be interesting to study the behavior of $\tilde{W}$ in the class of EMD theories \eqref{appaction} with generic non-relativistic geometries \eqref{appmetric}, and ask whether one could improve on the monotonicity properties we have identified.  Finally, recently
fixed-point annihilation and complex conformal field theories with complex RG flows have been discussed in holography for Einstein-scalar theories by \cite{Faedo:2019nxw}, who considered relativistic domain wall geometries. It would be interesting to extend the discussion to more general cases relevant for describing non-relativistic RG flows.
 We leave these questions to future work.

\acknowledgments\label{ACKNOWL}

We would like to thank Ioannis Papadimitriou and Matteo Baggioli for helpful conversations, and Elias Kiritsis for comments on the draft. We thank E. Blauvelt for initial collaboration.
S.C. is grateful to KITP for support while some of this research was conducted. 
The work of S.C. is supported in part by the National Science Foundation grant PHY-1915038. 
The work of L.L. is supported in part by the National Natural Science Foundation of China Grants No.11947302 and No.11991052.
The work of K.R. was supported in part by Lehigh University's REU program, through the National Science Foundation grant PHY-1852010. The work of Y.T. was funded by Lehigh University's Lee Fellowship.

\begin{appendices}

\section{Extremal geometry in Einstein-Scalar theory}\label{App:extreme}

In this section we put forth a non-trivial example that illustrates the fact that extremal geometries in Einstein-scalar theories are relativistic. 
We consider the following scalar potential,
\begin{eqnarray}
V=2 V_0\cosh[\delta\phi]+2 V_1\cosh[3\delta\phi]-6-2(V_0+V_1)\,,
\end{eqnarray}
which will be engineered to admit two extrema $(\phi=\phi_{UV}=0, \phi=\phi_{IR})$,
\begin{eqnarray}
V'(\phi=\phi_{UV})=0,\quad V'(\phi=\phi_{IR})=0\,,
\end{eqnarray}
located at the UV and IR values of the scalar. The domain wall solutions flow between two $AdS_4$ fixed points,
\begin{eqnarray}
ds^2_{AdS_4}=\frac{L^2}{u^2}{du^2}-\frac{u^2}{L^2}(-dt^2+dx^2+dy^2)\,,
\end{eqnarray}
with $L=(L_{IR}, L_{UV})$ parametrizing the size of the $AdS$ radius in the IR and UV, respectively. In particular,  the scalar field in the model rolls from a maximum of the potential at $\phi=0$ in the UV to a minimum in the IR at $\phi=\phi_{IR}$, with the RG flow traversing an intermediate regime between the two fixed points.

More precisely, expanding around the UV fixed point $\phi=0$, the potential $V$ reduces to
\begin{eqnarray}
V(\phi)=-6+\frac{1}{2}m_{UV}^2\phi^2+\mathcal{O}(\phi^4)\,,
\end{eqnarray}
from which we read off the UV AdS radius $L_{UV}$ and the scalar mass $m_{UV}^2$,
\begin{eqnarray}
L_{UV}=1,\quad\quad m_{UV}^2=2\delta^2 (V_0+9V_1)\,.
\end{eqnarray}
The requirement that $\phi=0$ is a maximum restricts $V_0+9V_1<0$. On the other hand, around the IR fix point the potential reads
\begin{eqnarray}
V(\phi)=-\frac{6}{L_{IR}^2}+\frac{1}{2}m_{IR}^2\phi^2+\mathcal{O}(\phi-\phi_{IR})^3\,,
\end{eqnarray}
with
\begin{eqnarray}
L_{IR}^2=-\frac{6}{V(\phi_{IR})},\quad\quad m_{IR}^2=2\delta^2 (V_0\cosh[\delta \phi_{IR}]+9V_1\cosh[3\delta\phi_{IR}])\,.
\end{eqnarray}

We now construct the full bulk solution which interpolates between the two $AdS_4$ fixed points. For the numerics we consider the following ansatz,
\begin{eqnarray}\label{RGmetric}
ds^2=\frac{1}{X(u)}du^2+X(u)\left(-\frac{Y(u)}{X(u)}dt^2+dx^2+dy^2\right),\quad \phi=\tilde{\phi}(u)\,.
\end{eqnarray}
To connect to our original coordinates~\eqref{appmetric}, one makes the following identification,
\begin{equation}
dr=\frac{du}{\sqrt{X(u)}},\quad e^{2A(r)}=X(u), \quad f(r)=\frac{Y(u)}{X(u)},\quad \phi(r)=\tilde{\phi}(u)\,.
\end{equation}
Notice that the \emph{extremal geometry is relativistic} when $f$ is a constant. 
Thus, our main task is to check whether the quantity $\frac{Y(u)}{X(u)}$ is independent of the radial coordinate $u$ or not. 
In our numerical analysis we first perform a series solution about the IR AdS fixed point, to sufficiently high order, and then identify irrelevant deformations triggered by the scalar perturbation. We then use this series solution to provide boundary conditions for the numerical evolution towards the UV fixed point.

\begin{figure}[ht!]
\begin{center}
\includegraphics[width=2.8in]{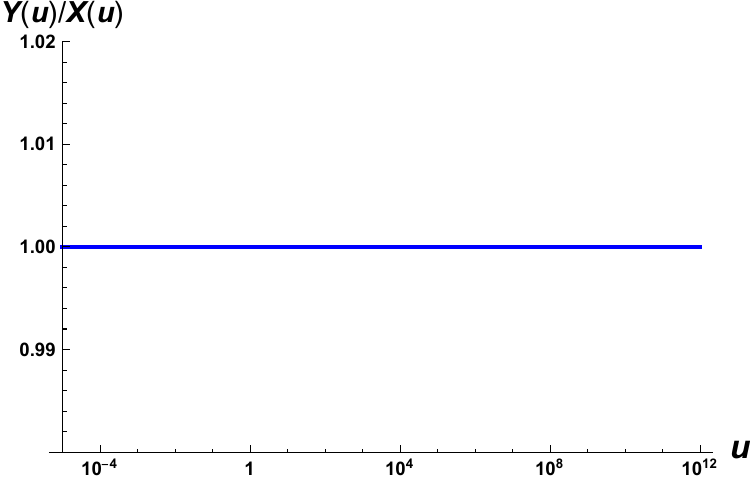}\quad
\includegraphics[width=2.8in]{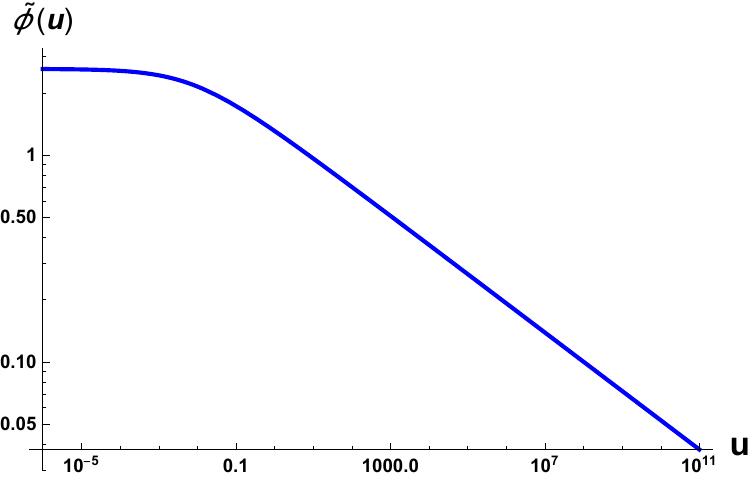}\quad
\includegraphics[width=2.8in]{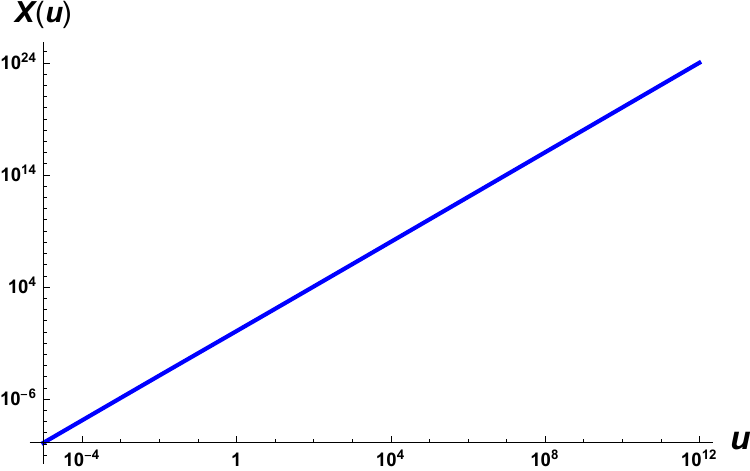}\quad
\includegraphics[width=2.8in]{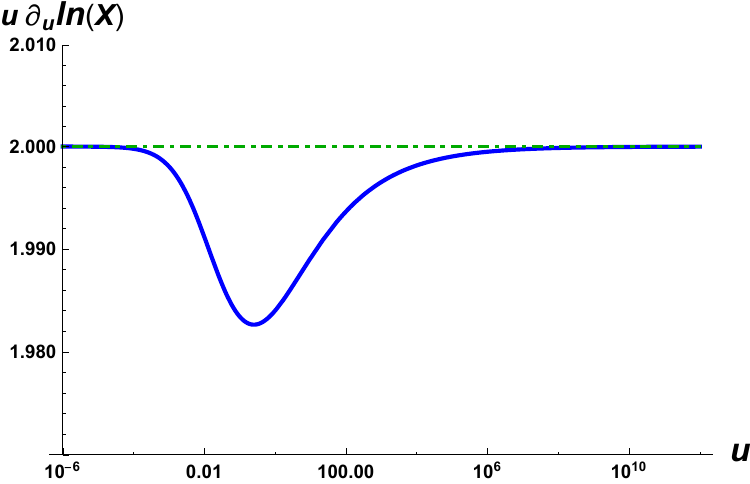}
\caption{Numerical solutions corresponding to the parameter choice \eqref{scaling1}. 
The solid blue curves denote the numerical solutions for the metric and scalar fields. The first plot shows that $X(u)$ and $Y(u)$ are identical, and thus the geometry is relativistic. The dot-dashed line in the fourth plot confirms that $X\sim u^{2}$ both in the UV and IR.}
\label{fig:RG}
\end{center}
\end{figure}

We consider two examples. In the first case, we choose the following parameters,
\begin{equation}\label{scaling1}
\delta=2/3,\quad V_0=-1/2,\quad V_1=1/200\,.
\end{equation}
The metric and scalar fields are shown by the blue curves in Figure~\ref{fig:RG}. 
Despite the presence of a non-trivial scalar field, it is manifest that two metric functions $X(u)$ and $Y(u)$ are identical, implying that the background geometry~\eqref{RGmetric} is relativistic. This confirms the observation that \emph{extremal geometries in Einstein-scalar theory are relativistic}
independently of which potential $V(\phi)$ one chooses.

\begin{figure}[h!]
\begin{center}
\includegraphics[width=2.8in]{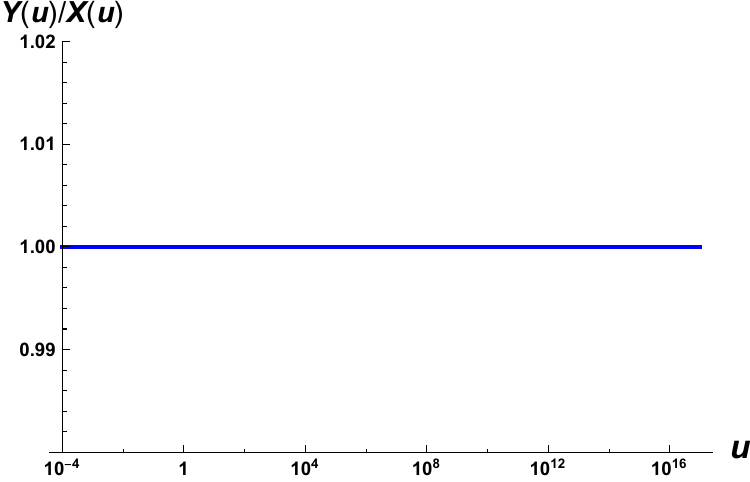}\quad
\includegraphics[width=2.8in]{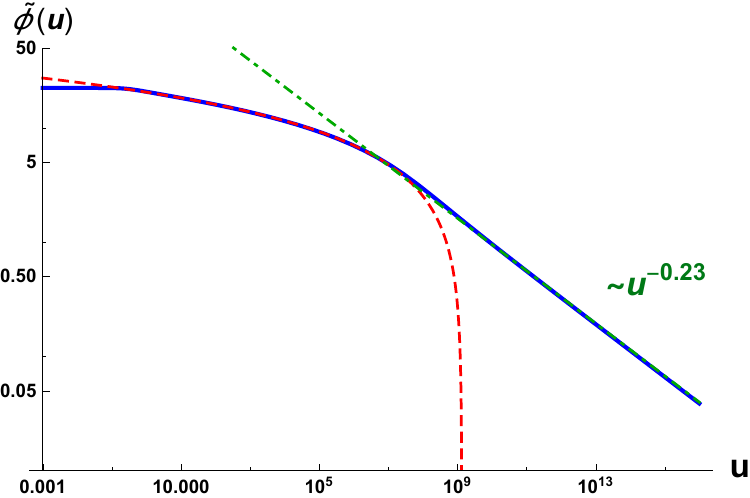}\quad
\includegraphics[width=2.8in]{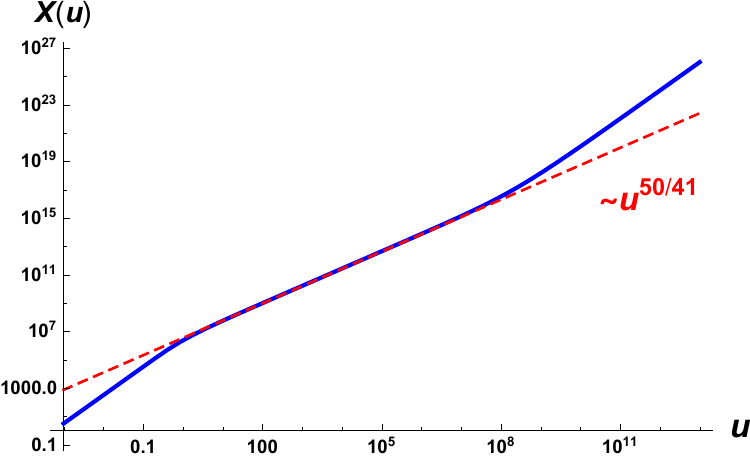}\quad
\includegraphics[width=2.8in]{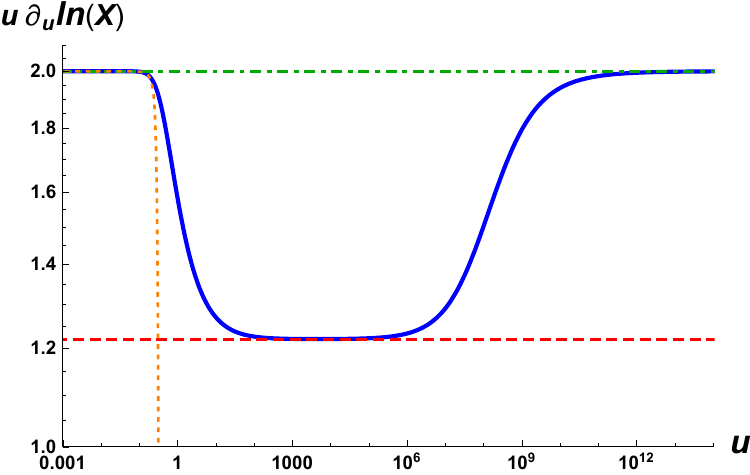}
\caption{Numerical solutions corresponding to the parameters~\eqref{scaling2}. The solid blue curves correspond to the numerical solutions for the metric and scalar. The first plot confirms that the geometry is relativistic. There is a clean hyperscaling violating intermediate regime shown by the dashed red lines in the second and third plots. In the intermediate region the metric function $X$ scales as $\sim u^{50/41}$, while the scalar field has the expected logarithmic behavior. The two straight lines in the fourth plot confirm that $X\sim u^{50/41}$ in the intermediate region and $X\sim u^{2}$ both in the UV and IR. The scalar in the UV goes to zero as $\sim u^{-0.23}$, as we turn on a non-trivial source for the dual scalar operator with the scaling dimension $\Delta\simeq3.23$. In the last plot we have added the IR series solution (dotted orange curve) to check agreement with our numerical solution in the IR region.}
\label{fig:RGscaling}
\end{center}
\end{figure}

For the second case, we choose the parameters to be
\begin{equation}\label{scaling2}
\delta=4/5,\quad V_0=-1/2,\quad V_1=5\times 10^{-17}\,,
\end{equation}
and the corresponding metric and scalar fields are shown in Figure~\ref{fig:RGscaling}. In this case one also finds a relativistic geometry, as expected. For the second set of parameters, there is an intermediate regime of hyperscaling violation, with the  hyperscaling violating parameter $\theta=-32/9$. In turn this implies a scaling in the function $X\sim u^{50/41}$ as well as in
the scalar, with $\tilde{\phi}\simeq c-\frac{40}{41} \text{ln}(u)$,  with the constant $c\simeq 20.5$ determined by fitting the numerical data. 
Both are confirmed by the log-log plot of $\tilde{\phi}(u)$ (second plot) and of $X(u)$ (third plot)  in Figure~\ref{fig:RGscaling}. The overlap between the red dashed curves  and the blue curves corresponding to the numerical solutions confirms the regime of hyperscaling violation in the intermediate region. Here $|V_1/V_0|\ll 1$ implies a large separation between the two AdS fixed points, and gives a clear intermediate scaling region.  It turns out that once $V_1$ is sufficiently small, the presence of an intermediate hyperscaling violation regime is quite generic\footnote{For a discussion of holographic RG flows with an intermediate scaling regime, see \emph{e.g.}~\cite{Bhattacharya:2014dea,Faulkner:2010gj}.}.

\section{Monotonicity for W in Einstein-scalar theory}\label{App:W}
It is well known that the superpotential $W$ is monotonic along the radial coordinate  for  vacuum (\emph{i.e.} Poincar\'e  invariant) solutions in Einstein-scalar theories, see also our discussion in subsection~\ref{SuperSubsection}. 
Probing this feature by considering non-vacuum states is important in order to test the robustness and consistency of these solutions. 
In this section we show that $W$ in Einstein-scalar theories is  no longer monotonic at finite temperature in general. 

We begin with the asymptotic expansions near the AdS boundary $r\rightarrow \infty$, which are schematically given by
\begin{eqnarray}
\phi(r)& = & \phi_s\, e^{-\Delta_{-} r/L}+\dots+\phi_v\, e^{-\Delta_{+} r/L}+\dots\,,\\
f(r)& = & 1- M e^{-d\,r/L}+\dots,\\
A(r)&=&\frac{r}{L}+\dots\,,
\end{eqnarray}
where $\Delta_{\pm}=(d\pm\sqrt{d^2 + 4m^2L^2})/2$ and $m^2$ is defined by~\eqref{expansionZV}. Using the conserved quantity $\mathcal{Q}$ in~\eqref{appQ}, one obtains
\begin{equation}
\mathcal{Q}=-2\kappa^2 T\, s=-\frac{d}{L}M\Rightarrow M=2\kappa^2 T\, s\frac{L}{d}\geqslant 0.
\end{equation}
In particular, at finite temperature one has $M>0$. 
Furthermore, here we only consider cases with $m^2$ beyond the AdS$_{d+1}$ BF bound, \emph{i.e.} $m^2>-\frac{d^2}{4L^2}$. 
Thus, we have $\Delta_{+}>d/2$ and $\Delta_{-}<d/2$.

When the holographic RG flow is driven by the expectation value of the scalar operator, \emph{i.e.} without the leading scalar source term in standard quantization, 
we obtain near the AdS boundary 
\begin{equation}
\frac{dW}{dr} = \dot\phi^2 - \frac{W W_A}{2(d-1)}=\phi_v^2\frac{\Delta_{+}^2}{L^2} e^{-2\Delta_+ r/L}+\dots-M\frac{d(d-1)}{L^2}e^{-d r/L}+\dots\,  ,
\end{equation}
Since $M>0$ at finite temperature and $\Delta_{+}>d/2$, one immediately finds that 
\begin{equation}
\frac{dW}{dr} = -M\frac{d(d-1)}{L^2}e^{-d r/L}+\dots<0\,  ,
\end{equation}
for the expectation-value-driven flow near the AdS boundary.

In contrast, for the RG flow driven by the scalar source, one has
\begin{equation}
\frac{dW}{dr} = \dot\phi^2 - \frac{W W_A}{2(d-1)}=\phi_s^2\frac{\Delta_{-}^2}{L^2} e^{-2\Delta_- r/L}+\dots-M\frac{d(d-1)}{L^2}e^{-d r/L}+\dots\,.
\end{equation}
Since $\Delta_{-}<d/2$, it is obvious that 
\begin{equation}
\frac{dW}{dr} =\phi_s^2\frac{\Delta_{-}^2}{L^2} e^{-2\Delta_- r/L}+\dots>0\,,
\end{equation}
near the UV boundary.

Let's now consider the behavior of $W$ near the horizon $r=r_h$. The near horizon expansion in the present coordinate system~\eqref{appmetric} reads\,\footnote{One might be more familiar with the coordinate system
\begin{equation}
ds^2 = \frac{du^2}{\hat{f}(u)} + e^{2\hat{A}(u)}(-\hat{f}(u) dt^2 + d\vec{x}^2) \, ,\quad \phi=\hat{\phi}(u), \quad A_\mu d x^\mu=\hat{A}_t(u) dt \, .
\end{equation}
Regularity demands the following IR expansion near the black hole horizon $u=u_h$: $\hat{\phi}(u) = \hat{\phi}_0+\hat{\phi}_1 (u-u_h)+\dots, \hat{f}(u) = \hat{f}_1 (u-u_h)+\dots$, and $\hat{A}(u) =\hat{A}_1 (u-u_h)+\dots$. To recover our original metric~\eqref{appmetric} we identify $dr=\frac{du}{\sqrt{\hat{f}(u)}}$, yielding $r-r_h\sim \sqrt{u-u_h}$ near the horizon.
}
\begin{eqnarray}
\phi(r)& = & \phi_0+\phi_1 (r-r_h)^{2}+\dots\,,\\
f(r)& = & f_1 (r-r_h)^{2}+\dots,\\
A(r)&=&A_1 (r-r_h)^{2}+\dots\,,
\end{eqnarray}
where $(\phi_0, \phi_1, f_1, A_1)$ are all constants. In particular, from subsections~\ref{sec:warp} and~\ref{sec:index} one finds $f_1>0$ and $A_1>0$ for black holes at finite temperature in  Einstein-scalar theories. 
Thus, one obtains
\begin{equation}
\frac{dW}{dr} =4\phi_1^2(r-r_h)^2(1+\dots)-4 A_1 (d-1)(1+\cdots)<0\,,
\end{equation}
near the black hole horizon. Combining the UV and IR analyses, we see that for the source-driven flow at finite temperature $\frac{dW}{dr}$ changes sign, and thus $W$ is not monotonic along the radial direction in this case.

\end{appendices}


\addcontentsline{toc}{section}{References}
\begin{thebibliography}{99}



\bibitem{Zamol}
A.~B.~Zamolodchikov, ``Irreversibility of the Flux of the Renormalization Group in a 2D
Field Theory,'' JETP Lett. 43, 730 (1986) [Pisma Zh. Eksp. Teor. Fiz. 43, 565 (1986)].

\bibitem{Komargodski:2011vj} 
  Z.~Komargodski and A.~Schwimmer,
  ``On Renormalization Group Flows in Four Dimensions,''
  JHEP {\bf 1112}, 099 (2011), arXiv:1107.3987 [hep-th].
 
\bibitem{Freedman:1999gp} 
  D.~Z.~Freedman, S.~S.~Gubser, K.~Pilch and N.~P.~Warner,
  ``Renormalization group flows from holography supersymmetry and a c theorem,''
  Adv.\ Theor.\ Math.\ Phys.\  {\bf 3}, 363 (1999)
  [hep-th/9904017].



\bibitem{deBoer:1999tgo}
J.~de Boer, E.~P.~Verlinde and H.~L.~Verlinde,
``On the holographic renormalization group,''
JHEP \textbf{08} (2000), 003
[arXiv:hep-th/9912012 [hep-th]].

\bibitem{Faulkner:2010jy}
T.~Faulkner, H.~Liu and M.~Rangamani,
``Integrating out geometry: Holographic Wilsonian RG and the membrane paradigm,''
JHEP \textbf{08} (2011), 051
[arXiv:1010.4036 [hep-th]].

\bibitem{Heemskerk:2010hk}
I.~Heemskerk and J.~Polchinski,
``Holographic and Wilsonian Renormalization Groups,''
JHEP \textbf{06} (2011), 031
[arXiv:1010.1264 [hep-th]].


\bibitem{Bhattacharya:2014dea}
J.~Bhattacharya, S.~Cremonini and B.~Gout\'eraux,
``Intermediate scalings in holographic RG flows and conductivities,''
JHEP \textbf{02} (2015), 035
[arXiv:1409.4797 [hep-th]].

\bibitem{Donos:2017ljs}
A.~Donos, J.~P.~Gauntlett, C.~Rosen and O.~Sosa-Rodriguez,
``Boomerang RG flows in M-theory with intermediate scaling,''
JHEP \textbf{07}, 128 (2017)
[arXiv:1705.03000 [hep-th]].

\bibitem{Donos:2017sba} 
  A.~Donos, J.~P.~Gauntlett, C.~Rosen and O.~Sosa-Rodriguez,
  ``Boomerang RG flows with intermediate conformal invariance,''
  JHEP {\bf 1804}, 017 (2018),
 [arXiv:1712.08017 [hep-th]].
  
  \bibitem{Hoyos:2020zeg} 
  C.~Hoyos, N.~Jokela, J.~M.~Penín and A.~V.~Ramallo,
  ``Holographic spontaneous anisotropy,''
  arXiv:2001.08218 [hep-th].
  


  
\bibitem{Rozali:2012es}
M.~Rozali, D.~Smyth, E.~Sorkin and J.~B.~Stang,
``Holographic Stripes,''
Phys. Rev. Lett. \textbf{110} (2013) no.20, 201603
[arXiv:1211.5600 [hep-th]].  

\bibitem{Withers:2014sja}
B.~Withers,
``Holographic Checkerboards,''
JHEP \textbf{09} (2014), 102
[arXiv:1407.1085 [hep-th]].
  
\bibitem{Cremonini:2017usb}
S.~Cremonini, L.~Li and J.~Ren,
``Intertwined Orders in Holography: Pair and Charge Density Waves,''
JHEP \textbf{08} (2017), 081
[arXiv:1705.05390 [hep-th]].
  
\bibitem{Cai:2017qdz}
R.~G.~Cai, L.~Li, Y.~Q.~Wang and J.~Zaanen,
``Intertwined Order and Holography: The Case of Parity Breaking Pair Density Waves,''
Phys. Rev. Lett. \textbf{119} (2017) no.18, 181601
[arXiv:1706.01470 [hep-th]].  
  
\bibitem{Andrade:2017ghg}
T.~Andrade, A.~Krikun, K.~Schalm and J.~Zaanen,
``Doping the holographic Mott insulator,''
Nature Phys. \textbf{14} (2018) no.10, 1049-1055
[arXiv:1710.05791 [hep-th]].  

\bibitem{Cremonini:2018xgj}
S.~Cremonini, L.~Li and J.~Ren,
``Holographic Fermions in Striped Phases,''
JHEP \textbf{12} (2018), 080
[arXiv:1807.11730 [hep-th]].

\bibitem{Ling:2019gjy}
Y.~Ling, P.~Liu and M.~H.~Wu,
``Holographic superconductor induced by charge density waves,''
[arXiv:1911.10368 [hep-th]].
  

\bibitem{Girardello:1998pd}
L.~Girardello, M.~Petrini, M.~Porrati and A.~Zaffaroni,
``Novel local CFT and exact results on perturbations of N=4 superYang Mills from AdS dynamics,''
JHEP \textbf{12}, 022 (1998)
[arXiv:hep-th/9810126 [hep-th]].
 


\bibitem{Papadimitriou:2007sj}
I.~Papadimitriou,
``Multi-Trace Deformations in AdS/CFT: Exploring the Vacuum Structure of the Deformed CFT,''
JHEP \textbf{05}, 075 (2007)
[arXiv:hep-th/0703152 [hep-th]].

\bibitem{Papadimitriou:2011qb}
I.~Papadimitriou,
``Holographic Renormalization of general dilaton-axion gravity,''
JHEP \textbf{08}, 119 (2011)
[arXiv:1106.4826 [hep-th]].

  
  
 \bibitem{Kiritsis:2016kog} 
  E.~Kiritsis, F.~Nitti and L.~Silva Pimenta,
  ``Exotic RG Flows from Holography,''
  Fortsch.\ Phys.\  {\bf 65}, no. 2, 1600120 (2017),
 arXiv:1611.05493 [hep-th].
 
\bibitem{Gursoy:2018umf}
U.~G\"ursoy, E.~Kiritsis, F.~Nitti and L.~Silva Pimenta,
``Exotic holographic RG flows at finite temperature,''
JHEP \textbf{10}, 173 (2018)
[arXiv:1805.01769 [hep-th]].

\bibitem{Bea:2018whf}
Y.~Bea and D.~Mateos,
``Heating up Exotic RG Flows with Holography,''
JHEP \textbf{08}, 034 (2018)
[arXiv:1805.01806 [hep-th]].

\bibitem{Kiritsis:2012ma}
E.~Kiritsis and V.~Niarchos,
``The holographic quantum effective potential at finite temperature and density,''
JHEP \textbf{08}, 164 (2012)
[arXiv:1205.6205 [hep-th]].
  
  
\bibitem{Lindgren:2015lia} 
  J.~Lindgren, I.~Papadimitriou, A.~Taliotis and J.~Vanhoof,
  ``Holographic Hall conductivities from dyonic backgrounds,''
  JHEP {\bf 1507}, 094 (2015),
 [arXiv:1505.04131 [hep-th]].
  
\bibitem{Cremonini:2013ipa} 
  S.~Cremonini and X.~Dong,  ``Constraints on renormalization group flows from holographic entanglement entropy,''
  Phys.\ Rev.\ D {\bf 89}, no. 6, 065041 (2014), 
  [arXiv:1311.3307 [hep-th]].



\bibitem{Casini:2004bw} 
  H.~Casini and M.~Huerta,
  ``A Finite entanglement entropy and the c-theorem,''
  Phys.\ Lett.\ B {\bf 600}, 142 (2004)
  [hep-th/0405111].  
  
  \bibitem{Myers:2010tj} 
  R.~C.~Myers and A.~Sinha,
  ``Holographic c-theorems in arbitrary dimensions,''
  JHEP {\bf 1101}, 125 (2011)
  [arXiv:1011.5819 [hep-th]].
  
  \bibitem{Casini:2012ei} 
  H.~Casini and M.~Huerta,
  ``On the RG running of the entanglement entropy of a circle,''
  Phys.\ Rev.\ D {\bf 85}, 125016 (2012)
  [arXiv:1202.5650 [hep-th]].
  
  \bibitem{Banerjee:2015coc}
S.~Banerjee and P.~Paul,
``Black Hole Singularity, Generalized (Holographic) $c$-Theorem and Entanglement Negativity,''
JHEP \textbf{02}, 043 (2017)
[arXiv:1512.02232 [hep-th]].
  
  \bibitem{Chu:2019uoh}
C.~S.~Chu and D.~Giataganas,
``$c$-Theorem for Anisotropic RG Flows from Holographic Entanglement Entropy,''
Phys. Rev. D \textbf{101}, no.4, 046007 (2020)
[arXiv:1906.09620 [hep-th]].
  
\bibitem{Swingle:2013zla} 
  B.~Swingle,
  ``Entanglement does not generally decrease under renormalization,''
  J.\ Stat.\ Mech.\  {\bf 1410}, no. 10, P10041 (2014)
  [arXiv:1307.8117 [cond-mat.stat-mech]].  


\bibitem{Kiritsis:2015hoa} 
  E.~Kiritsis and L.~Li,
  ``Holographic Competition of Phases and Superconductivity,''
  JHEP {\bf 1601}, 147 (2016)
  [arXiv:1510.00020 [cond-mat.str-el]].
  
\bibitem{Li:2020spf}
L.~Li,
``On Thermodynamics of AdS Black Holes with Scalar Hair,''
[arXiv:2008.05597 [gr-qc]].

\bibitem{Bousso:1999xy} 
  R.~Bousso,
  ``A Covariant entropy conjecture,''
  JHEP {\bf 9907}, 004 (1999)
  [hep-th/9905177].


\bibitem{Sahakian:1999bd} 
  V.~Sahakian,
  ``Holography, a covariant c function, and the geometry of the renormalization group,''
  Phys.\ Rev.\ D {\bf 62}, 126011 (2000),
  [hep-th/9910099].
  
\bibitem{Kolekar:2018chf}
K.~S.~Kolekar and K.~Narayan,
``On AdS$_{2}$ holography from redux, renormalization group flows and c-functions,''
JHEP \textbf{02}, 039 (2019)
[arXiv:1810.12528 [hep-th]].
 
 
 
\bibitem{Gubser:2009gp} 
  S.~S.~Gubser, S.~S.~Pufu and F.~D.~Rocha,
  ``Quantum critical superconductors in string theory and M-theory,''
  Phys.\ Lett.\ B {\bf 683}, 201 (2010),
  [arXiv:0908.0011 [hep-th]].
  

\bibitem{Kiritsis:1999tx}
E.~Kiritsis,
``Supergravity, D-brane probes and thermal superYang-Mills: A Comparison,''
JHEP \textbf{10}, 010 (1999)
[arXiv:hep-th/9906206 [hep-th]].


\bibitem{Gauntlett:2018vhk}
J.~P.~Gauntlett and C.~Rosen,
``Susy Q and spatially modulated deformations of ABJM theory,''
JHEP \textbf{10}, 066 (2018)
[arXiv:1808.02488 [hep-th]].

\bibitem{Arav:2018njv}
I.~Arav, J.~P.~Gauntlett, M.~Roberts and C.~Rosen,
``Spatially modulated and supersymmetric deformations of ABJM theory,''
JHEP \textbf{04}, 099 (2019)
[arXiv:1812.11159 [hep-th]].


\bibitem{Myers:2012ed} 
  R.~C.~Myers and A.~Singh,
  ``Comments on Holographic Entanglement Entropy and RG Flows,''
  JHEP {\bf 1204}, 122 (2012),
 [arXiv:1202.2068 [hep-th]].


  

  
\bibitem{Cvetic:1999xp}
M.~Cvetic, M.~Duff, P.~Hoxha, J.~T.~Liu, H.~Lu, J.~Lu, R.~Martinez-Acosta, C.~Pope, H.~Sati and T.~A.~Tran,
``Embedding AdS black holes in ten-dimensions and eleven-dimensions,''
Nucl. Phys. B \textbf{558}, 96-126 (1999)
[arXiv:hep-th/9903214 [hep-th]].

\bibitem{Kiritsis:2015oxa}
E.~Kiritsis and J.~Ren,
``On Holographic Insulators and Supersolids,''
JHEP \textbf{09} (2015), 168
[arXiv:1503.03481 [hep-th]].  


  

\bibitem{DeWolfe:2012uv}
O.~DeWolfe, S.~S.~Gubser and C.~Rosen,
``Fermi surfaces in N=4 Super-Yang-Mills theory,''
Phys. Rev. D \textbf{86}, 106002 (2012)
[arXiv:1207.3352 [hep-th]].



\bibitem{Faedo:2019nxw}
A.~F.~Faedo, C.~Hoyos, D.~Mateos and J.~G.~Subils,
``Holographic Complex Conformal Field Theories,''
Phys. Rev. Lett. \textbf{124} (2020) no.16, 161601
[arXiv:1909.04008 [hep-th]].



\bibitem{Faulkner:2010gj}
T.~Faulkner, G.~T.~Horowitz and M.~M.~Roberts,
``Holographic quantum criticality from multi-trace deformations,''
JHEP \textbf{04} (2011), 051
[arXiv:1008.1581 [hep-th]].







\end{thebibliography}
\end{document}